\documentclass{aa}

\usepackage{graphicx}
\usepackage{txfonts}
\usepackage{textgreek}
\usepackage{textcomp}
\usepackage{tablefootnote}
\usepackage{natbib}
\bibpunct{(}{)}{;}{a}{}{,} % to follow the A&A style
\begin{document}

   \title{VLBI Celestial and Terrestrial Reference Frames VIE2022b}

% VLBI Celestial and Terrestrial Reference Frames VIE2020 and VIE2022b

   \author{H. Kr\'asn\'a \inst{1}
          \and
          L. Baldreich \inst{1}
          \and
          J. B\"ohm \inst{1}
          \and
          S. B\"ohm \inst{1}
          \and
          J. Gruber \inst{1}
          \and
          A. Hellerschmied \inst{2}
          \and
           F. Jaron \inst{1}
          \and
          L. Kern \inst{1}
          \and
          D. Mayer \inst{2}
          \and
          A. Nothnagel \inst{1}
          \and
          O. Panzenb\"ock \inst{1}
          \and
          H. Wolf \inst{1}
          }

   \institute{ Technische Universit\"at Wien (TU Wien), Department of Geodesy and Geoinformation,  Wiedner Hauptstraße 8-10, 1040 Vienna, Austria\\
              \email{hana.krasna@tuwien.ac.at}
              \and
              Federal Office of Metrology and Surveying (BEV), Schiffamtsgasse 1-3, 1020 Vienna, Austria
             }

   \date{Received ; accepted }

% \abstract{}{}{}{}{}
% 5 {} token are mandatory

  \abstract
   {We introduce the computation of global reference frames from Very Long Baseline Interferometry (VLBI) observations at the Vienna International VLBI Service for Geodesy and Astrometry (IVS) Analysis Center (VIE) in detail. We focus on the celestial and terrestrial frames from our two latest solutions VIE2020 and VIE2022b.}
   {The current International Celestial and Terrestrial Reference Frames, ICRF3 and ITRF2020, include VLBI observations until spring 2018 and December 2020, respectively. We provide terrestrial and celestial reference frames including VLBI sessions until June 2022 organized by the IVS.}
   {Vienna terrestrial and celestial reference frames are computed in a common least squares adjustment of geodetic and astrometric VLBI observations with the Vienna VLBI and Satellite Software (VieVS).}
   {We provide high-quality celestial and terrestrial reference frames computed from 24-hour IVS observing sessions. The CRF provides positions of 5407 radio sources. In particular, positions of sources with few observations at the time of the ICRF3 calculation could be improved. The frame also includes positions of 870 new radio sources, which are not included in ICRF3.
   The additional observations beyond the data used for ITRF2020 provide a more reliable estimation of positions and linear velocities of newly established  VLBI Global Observing System (VGOS) telescopes.}
   {}


   \maketitle
%
%________________________________________________________________

\section{Introduction}

Geodetic and astrometric Very Long Baseline Interferometry (VLBI) observations have been carried out since 1979. Since 1999, the VLBI sessions -- from scheduling to analyses -- have been organized by the International VLBI Service for Geodesy and Astrometry \citep[IVS;][]{Nothnagel2017}.
The Vienna IVS Analysis Center (VIE), jointly operated by TU Wien and the Federal Office of Metrology and Surveying (BEV) in Austria, is one of the eight international institutions worldwide which actively contributed to studies needed for the latest update of the International Celestial Reference Frame (ICRF) generating one of the prototype realizations of the current ICRF3~\citep{Charlot20}, which was adopted by the International Astronomical Union in August 2018.

Furthermore, VIE, as one of the IVS Analysis Centers, participated in recent international efforts of calculating an updated realization of the International Terrestrial Reference System (ITRS). We analyzed a complete set of individual VLBI observing sessions and provided pre-reduced normal equation systems (NEQ) for combination with results of other IVS Analysis Centers at the IVS Combination Center \citep{Hellmers2022}. The final International Terrestrial Reference Frame (ITRF) originates from a further combination of reprocessed solutions from four space geodetic techniques (Doppler Orbitography by Radiopositioning Integrated on Satellite (DORIS), Global Navigation Satellite Systems (GNSS), Satellite Laser Ranging (SLR), and VLBI). This final combination allows to overcome the weaknesses of individual techniques taking advantage of the strength of a combined solution. The current version of the International Terrestrial Reference Frame, the ITRF2020 \citep{Altamimi22}, was published in April 2022.

Besides our support of the international efforts towards conventional reference frame realizations, we generate our own global VLBI reference frames. The celestial and terrestrial reference frames, together with the connecting Earth Orientation Parameters (EOP), are created in a common least squares adjustment of VLBI observations using the Vienna VLBI and Satellite Software version 3.3 \citep[VieVS;][]{Boehm18}. In this paper, we introduce our global reference frame solution VIE2020 generated with VLBI observing sessions as provided for the ITRF2020 computations. Furthermore, we also focus on our recent global solution VIE2022b, which involves VLBI experiments released after the ITRF2020 data deadline. In total, this solution includes VLBI data from additional 18 months compared to ITRF2020. In Section~\ref{sec_data}, we describe the VLBI data in detail and give information about the a priori models and parametrization of the solutions. Extensive descriptions of the estimated terrestrial and celestial reference frames, including comparisons to the most recent international reference frames, are given in Sections~\ref{sec_trf} and~\ref{sec_crf}, respectively.

\section{Data and analysis settings}
\label{sec_data}
We analyze VLBI group delays, which are fundamental observables of geodetic and global astrometric VLBI. In the IVS operation scheme \citep{Nothnagel2017}, after correlation, fringe fitting and pre-processing, the group delays are provided in so-called databases in the IVS vgosDB format via IVS Data Centers for Level 2 data analysis. Databases of version 4 include group delays at X-band (8.4~GHz) with ambiguities resolved and the ionospheric delay calibrated with a linear combination with simultaneous S-band (2.3~GHz) measurements.
\begin{table}
\caption{Overview of solutions described by this publication.}
\label{tab:solutions}       % Give a unique label
\begin{tabular}{llll}
\hline\noalign{\smallskip}
 &  & no. of  & no. of   \\
solution & data span &  sessions &  observations  \\
\noalign{\smallskip}\hline\noalign{\smallskip}
\hline
VIE2020 & 1979.5 - 2021.0 & 6786 & $20.0\cdot10^6$\\
VIE2020-sx & 1979.5 - 2021.0 &6748 &  $19.7\cdot10^6$\\
VIE2022b & 1979.5 - 2022.5 & 7148 &  $22.4\cdot10^6$\\
VIE2022b-sx & 1979.5 - 2022.5 &  7058&  $21.6\cdot10^6$\\
\noalign{\smallskip}\hline
\end{tabular}
\end{table}
\begin{table*}
\caption{A priori models used in the VIE2020 and VIE2022b solutions. The asterisk symbol denotes parameters modeled differently from our ITRF2020 contribution.}
\label{tab:statmodel}       % Give a unique label
\begin{tabular}{ll}
\hline\noalign{\smallskip}
 A priori modeling&  \\
\noalign{\smallskip}\hline\noalign{\smallskip}
\hline
\textit{Station a priori position}&\\
TRF with post-seismic deformation & ITRF2014 model \citep{Altamimi16} for VIE2020\\
& ITRF2020 model \citep{Altamimi22} for VIE2022b\\
\textit{Station displacement}&\\
solid Earth tides & IERS Conventions 2010 \citep{iers10}\\
rotational deformation due to polar motion & secular polar motion (updated IERS Conventions 2010, \href{https://iers-conventions.obspm.fr/content/chapter7/icc7.pdf}{version 2018-02-01}) \\
ocean pole tide loading & IERS Conventions 2010 \\
ocean tidal loading  & TPXO72 \citep{Egbert2002}\\
atmospheric tidal and non-tidal loading & APL-VIENNA \citep{Wijaya2013}\\
\noalign{\smallskip}\hline\noalign{\smallskip}
\textit{Earth orientation parameters} & \\
$^*$daily EOP & IERS Bulletin A, \href{https://maia.usno.navy.mil/ser7/finals2000A.all}{finals2000A.all} \\
subdaily EOP model & \citet{Desai2016} (ocean tides + libration; linear interpolation)\\
tidal UT variations & IERS Conventions 2010, UT1S all constituents\\
precession/nutation model & IAU 2006/2000A \citep{Mathews2002, Capitaine2003}  \\
\noalign{\smallskip}\hline\noalign{\smallskip}
\textit{Troposphere}& \\
hydrostatic delay& in situ pressure \citep{Saastamoinen1972}\\
hydrostatic mapping function & VMF3 \citep{Landskron2018}\\
hydrostatic gradients &  DAO gradients \citep{MacMillan1997}\\
$^*$wet mapping function&  VMF3  \citep{Landskron2018}\\
\noalign{\smallskip}\hline\noalign{\smallskip}
\textit{VLBI specific effects}&\\
thermal antenna deformation  & \citet{Nothnagel2009} with in situ temperature and GPT3 as backup\\
antenna axis offsets & \citet{Nothnagel2009}, antenna-info.txt version 2020-04-23\\
station eccentricities & \href{https://ivscc.gsfc.nasa.gov/IVS_AC/apriori/ECCDAT_v2019Dec19.ecc}{ECCDAT\_v2019Dec19.ecc}
\\
gravitational antenna deformation & \citet{Nothnagel2014}\\
CRF & ICRF3 \citep{Charlot20} \\
galactic aberration & modeled with  $\alpha_{GC} = 266.4^{\circ}, \delta_{GC} =-28.94^{\circ}, A_G = 5.8~$\textmu as/yr, epoch 2015.0\\
\noalign{\smallskip}\hline
\end{tabular}
\end{table*}
\begin{table*}
\caption{Parametrization of the VIE2020 and VIE2022b solutions. The asterisk symbol denotes a different analysis setting compared with our ITRF2020 contribution.}
\label{tab:solparam}       % Give a unique label
\begin{tabular}{ll}
\hline\noalign{\smallskip}
 Parametrization options&   \\
\noalign{\smallskip}\hline\noalign{\smallskip}
\hline
clocks &piece-wise linear offsets (PWLO) with time interval 1 hour with relative constraints 43 ps (1.3~cm) \\
&  between offsets, one rate and  quadratic term \\
baseline clock offsets  & offset without constraints at selected baselines\\
zenith wet delay & PWLO with time interval 30 min with relative constraints 50~ps (1.5~cm) between offsets\\
tropo. gradients & PWLO with time interval 3 hours with relative constraints 0.5~mm between offsets\\
$^*$ERP  & PWLO with time interval 24 hours with relative constraints 10~mas \\
& fixed for networks with less than four telescopes\\
$^*$celestial pole offsets & PWLO with time interval 24 hours with relative constraints 0.1~\textmu as \\
& fixed for networks with less than four telescopes\\
$^*$weighting & database weights (1/uncertainties)  + elevation dependent weighting\\
\hline\noalign{\smallskip}
Datum definition&   \\
\noalign{\smallskip}\hline\noalign{\smallskip}
CRF & NNR w.r.t. ICRF3 defining sources except 0700-465, 0809-493 \\
TRF & NNT/NNR on coordinates and velocities w.r.t. ITRF2014 on 21 stations for VIE2020\\
    & NNT/NNR on coordinates and velocities w.r.t. ITRF2020 on 21 stations for VIE2022b\\
\noalign{\smallskip}\hline
\end{tabular}
\end{table*}
\begin{figure*}
   \centering
   \includegraphics[width=\hsize]{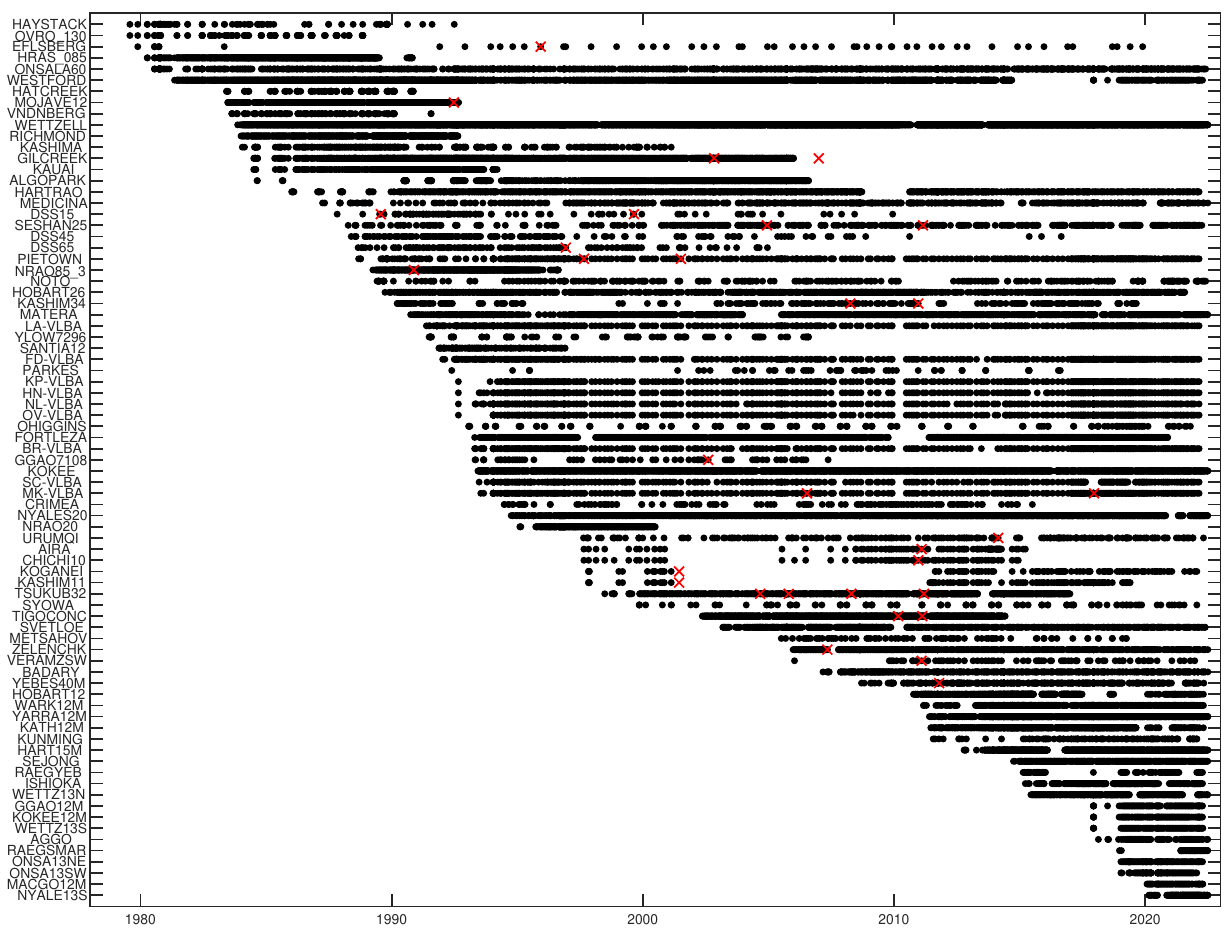}
      \caption{Antenna participation in VLBI experiments in solution VIE2022b. Red crosses depict the placement of position breaks. Only antennas with more than 50~experiments are plotted.
              }
         \label{FigTRFantennas50}
\end{figure*}
The solution VIE2020 is based on VLBI experiments observed at S/X frequencies, which were analyzed for the ITRF2020 contribution\footnote{\href{ https://ivscc.gsfc.nasa.gov/IVS_AC/ITRF2020/itrf2020_sx_sessionTable_v2021Feb10.txt}{itrf2020\_sx\_sessionTable\_v2021Feb10.txt}} by the IVS. In addition, 467~S/X experiments, scheduled mainly for astrometric purposes to strengthen the ICRF in the Southern Hemisphere, were added to the VIE2020 solution.

In recent years the geodetic VLBI technique has been undergoing a transition phase from a legacy S/X era to a novel VLBI Global Observing System \citep[VGOS; ][]{Petrachenko2009}. The VGOS system is based on the so-called broadband delay, which uses several frequency bands in the range from 2.5~GHz to 14~GHz. It had been decided that also the VGOS observations\footnote{\href{ https://ivscc.gsfc.nasa.gov/IVS_AC/ITRF2020/itrf2020_vgos_sessionTable_v2021Feb10.txt}{itrf2020\_vgos\_sessionTable\_v2021Feb10.txt}} were included in the ITRF2020 solution. This enables a consistent estimation of the coordinates for newly built VGOS telescopes within the international terrestrial reference frame. On the other hand, if a global solution is computed with a primary focus on the celestial reference frame, the mixture of observations at different frequencies is not desirable. Therefore, we provide two solutions, with and without VGOS sessions, denoted further as VIE2020 and VIE2020-sx, respectively. Finally, both datasets have been extended by sessions which arrived after the ITRF2020 data deadline, i.e., from January 2021 until June 2022. The solutions are denoted as VIE2022b and VIE2022b-sx (see Table~\ref{tab:solutions}). There are 7148 VLBI sessions in the VIE2022b solution providing 22.4~million observations in total. The solution VIE2022b-sx is built by 21.6~million observations which makes a difference of 0.8~million observations generated by 90 24-hour VGOS experiments.

In Table~\ref{tab:statmodel}, we provide an overview of the a priori models used for the calculation of the VLBI theoretical delays in the VIE2020 and VIE2022b solutions. The only difference is the choice of the a priori terrestrial reference frame: VIE2020 is modeled with ITRF2014 whereas VIE2022b is based on the recently available ITRF2020. The asterisk symbol in Table~\ref{tab:statmodel} highlights the application of a different model compared to our submission for the ITRF2020. At this point, we want to stress the importance of a consistent use of tropospheric mapping functions, more specifically, the same version of the Vienna mapping functions (VMF). Mixing VMF3 \citep{Landskron2018} and VMF1 \citep{Boehm2006} for modeling the hydrostatic and wet part of the tropospheric delays, respectively, leads to an estimated baseline length that is on average 2~mm shorter compared to using VMF1 or VMF3 for both constituents. In-depth analyses on that matter are ongoing.

In contrast to our ITRF2020 submission, the stochastic model, consisting of the standard deviations derived from the fringe fitting process and elevation-dependent observational noise $\sigma_{obs}$ \citep[Eq.~(\ref{eq_diag_elevdep});][] {Gipson08}, is applied as a diagonal covariance matrix in the adjustment:
\begin{equation}
\sigma^2_{obs} = \sigma^2_{0} +   \sigma^2_{i} +  \sigma^2_{j} = \sigma^2_{0} + \bigg(\frac{x_i}{\sin(\varepsilon_i)}\bigg)^2  + \bigg(\frac{x_j}{\sin(\varepsilon_j)}\bigg)^2 ,
\label{eq_diag_elevdep}
\end{equation}
where $\sigma_{0}$ consists of a measurement noise $\sigma_{m}$ together with an ionospheric delay formal error $\sigma_{ion}$: $\sigma_{0}^2 = \sigma_{m}^2 + \sigma_{ion}^2$. The elevation-dependent noise terms for stations $i$ and $j$, $\sigma_{i}$ and $\sigma_{j}$, are computed with a constant $x$ = 6~ps for each station divided by the sine of the associated elevation angle $\varepsilon$. Hence, observations at lower averaged elevations obtain a lower weight in the least squares adjustment.

In the so-called first solution, a pre-analysis of the VLBI experiments with a basic parametrization is carried out to check for possible problematic behavior of the data. In the majority of cases, this can be solved with an individual single session parametrization, e.g., the definition of clock breaks, exclusion of cable calibration signals at stations, removal of large outliers and removal of measurements at individual antennas or baselines, if unavoidable. This configuration with even more possible options for each VLBI experiment is stored in the VieVS-internal OPT-files\footnote{\href{https://github.com/TUW-VieVS/VLBI_OPT}{https://github.com/TUW-VieVS/VLBI\_OPT}} which we provide to the public.

As a next step, the parametrization of the so-called main solution is configured (see Table~\ref{tab:solparam}). In the software package VieVS the parametrization of the time-variable unknowns is realized with piece-wise linear offsets (PWLO) with certain time intervals and relative constraints between the offsets. The single session least-squares adjustment is also important for identifying baseline-dependent clock offsets \citep[BCO; ][]{Krasna2021} in the network. The baselines with an estimated clock offset larger than three times its formal error are listed in the OPT-files for estimation in the final solution.

Next, session-wise normal equation systems (NEQ) are prepared for the final global adjustment. After the reduction, these NEQ contain only parameters constant over several VLBI experiments, which will be estimated globally. The remaining parameters, which are time-dependent and do not profit from long time series of observation, are squeezed out from the NEQ. In our solutions, we estimate session-wise (baseline-dependent) clock parameters, zenith wet delays, tropospheric gradients in north and east direction at individual stations, and coordinates of telescopes which observed only occasionally within our dataset. The Earth rotation parameters (ERP, i.e., polar motion and UT1-UTC) are estimated session-wise for networks with more than three telescopes. For the small networks of three stations or single baselines, all five Earth orientation parameters (EOP) are fixed to a priori International Earth Rotation and Reference Systems Service (IERS) Bulletin A values issued by the IERS Rapid Service/Prediction Center at the U.S. Naval Observatory (USNO).

The adjustment of the ensemble of sessions is applied to a normal equation system which results from stacking the individual NEQ according to the theorem of stacking pre-reduced normal equation systems by Helmert. The set of global parameters determined by the common inversion of the stacked NEQ then only contains coordinates and velocities of telescopes, amplitudes of annual and semi-annual displacements in the station height of selected stations, and positions of radio sources.

\section{Vienna terrestrial reference frame}
\label{sec_trf}

Producing a new global terrestrial reference frame (TRF) requires detailed information about the underlying dataset and in particular about the observation time of telescopes taking part in the observation schedules. A traditional TRF catalog consists of station coordinates at a given epoch and the linear velocity. The velocity determination requires a sufficiently long observation time series of the telescope in order to enable a confident approximation of the antenna linear movement caused predominantly by the underlying plate tectonics. For this reason, we set a lower limit of 15 VLBI experiments including the particular station and a minimum time span of five years between the first and last experiment. For telescopes, which do not fulfill these conditions, the position is estimated as time series from individual sessions which is averaged a posteriori to the final single position. Exceptions to this rule are telescopes at co-location sites with older VLBI telescopes. In this case, the corrections to the a priori velocity for both (all) telescopes are tied together with a relative constraint which forces the velocity corrections at the specified telescopes to be identical. The approach requires that the same a priori velocity is used for all involved telescopes. The definition of interval breaks because of discontinuous station movements, as well as the modeling of the non-linear motion caused by post seismic deformation, is taken from the underlying a priori catalog, i.e., ITRF2014 and ITRF2020 for VIE2020 and VIE2022b solutions, respectively. Table~\ref{tab:velocties} summarizes the pairs (groups) of telescopes tied together in solution VIE2022b.
\begin{table}
\tiny
\caption{Co-located VLBI telescopes in VIE2022b where velocity ties are applied.}
\label{tab:velocties}       % Give a unique label
\begin{tabular}{l}
\hline\noalign{\smallskip}
telescopes - location \\
\noalign{\smallskip}\hline\noalign{\smallskip}
\hline
DSS15,     DSS13     - CA, USA\\
DSS65,     DSS65A,    ROBLED32 - Spain \\
FD-VLBA,   MACGO12M - TX, USA\\
FORTORDS,  FORT\_ORD - CA, USA \\
GGAO7108,  GORF7102,  GGAO12M  - MD, USA \\
HARTRAO,   HART15M  - South Africa \\
HOBART26,  HOBART12 - TAS, Australia\\
HRAS\_085,  FTD\_7900 - TX, USA\\
KASHIM34,  KASHIM11,  KASHIMA  - Japan \\
KAUAI, KOKEE,  KOKEE12M - HI, USA \\
METSAHOV,  METSHOVI - Finland\\
MIZNAO10,   VERAMZSW  - Japan \\
MOJAVE12,  MOJ\_7288 - CA, USA\\
NRAO20,    NRAO\_140,  NRAO85\_1, NRAO85\_3  - WV, USA\\
NYALES20,  NYALE13S - Norway \\
ONSALA60,   ONSA13NE,  ONSA13SW - Sweden\\
OV-VLBA, OVRO\_130,  OVR\_7853 - CA, USA\\
PIETOWN,   VLA-N8   - NM, USA\\
RICHMOND,  MIAMI20 - FL, USA \\
SVETLOE,   SVERT13V - Russia\\
WETTZELL,  TIGOWTZL,  WETTZ13N, WETTZ13S - Germany \\
YEBES,     YEBES40M,  RAEGYEB - Spain\\
YLOW7296,  YELLOWKN - Canada \\
\noalign{\smallskip}\hline
\end{tabular}
\end{table}

\begin{table*}
\caption{Annual and semi-annual signal (described with amplitude $A$ and phase $\phi$) in station height of new VGOS antennas and of appertaining legacy antennas at the common sites observing in more than 50 sessions in VIE2022b distributed uniformly over the year. For comparison, the harmonic signal in height from ITRF2020 for the entire site is noted in the respective last rows.}
\label{tab:harmpos}       % Give a unique label
\begin{tabular}{l|r|r|r|r|r|r|r}
\hline\noalign{\smallskip}
antenna & TRF& $A$ [mm] & $\phi$ [$^{\circ}$] & $A$ [mm] & $\phi$ [$^{\circ}$] & no. of sessions & data span [yr:doy]\\
 & &\multicolumn{2}{|c|}{annual signal} & \multicolumn{2}{|c|}{semi-annual signal} & & \\
\noalign{\smallskip}\hline\noalign{\smallskip}
\hline
GGAO12M  & VIE2022b &  2.7 $\pm$  0.3 &    240 $\pm$      05 &          0.3 $\pm$  0.3 &    130 $\pm$     47        & 85 & 2017:338 - 2022:098 \\
GGAO7108  & VIE2022b &   7.9 $\pm$  3.3 &    187 $\pm$     21 &          2.0 $\pm$  3.3 &    149 $\pm$     86        & 65 &   1993:118 - 2002:219  \\
         & ITRF2020 &    3.1 $\pm$    0.2  &    226 $\pm$ 04  &    0.2 $\pm$    0.2  &    105 $\pm$ 75  &  &    \\
\noalign{\smallskip}\hline\noalign{\smallskip}
KOKEE12M  & VIE2022b &   2.1 $\pm$  0.5 &    123 $\pm$     13 &          2.3 $\pm$  0.5 &    145 $\pm$     12      &    86    &    2017:338 - 2022:098  \\
KOKEE     & VIE2022b &   2.4 $\pm$  0.2 &    172 $\pm$      05 &          0.6 $\pm$  0.2 &     37 $\pm$     20      &    2790  &     1993:160 - 2022:175  \\
         & ITRF2020 &    1.1 $\pm$    0.3  &    160 $\pm$ 17  &    0.4 $\pm$    0.3  &    115 $\pm$ 47  &  &    \\
\noalign{\smallskip}\hline\noalign{\smallskip}
MACGO12M  & VIE2022b &   2.6 $\pm$  0.3 &    325 $\pm$      07 &          0.9 $\pm$  0.3 &     30 $\pm$     22           &  56   &     2020:022 - 2022:102   \\
FD-VLBA   & VIE2022b &   1.3 $\pm$  0.2 &    220 $\pm$     11 &          1.0 $\pm$  0.2 &    156 $\pm$     14            &   405  &      1992:014 - 2022:048  \\
HRAS\_085  & VIE2022b &   3.1 $\pm$  2.6 &    177 $\pm$     40 &          3.7 $\pm$  2.4 &     39 $\pm$     37            &   729   &    1980:103 - 1990:303   \\
         & ITRF2020 &    3.1 $\pm$    0.4  &    176 $\pm$ 06  &    0.5 $\pm$    0.3  &    151 $\pm$ 39  &  &    \\
\noalign{\smallskip}\hline\noalign{\smallskip}
NYALE13S  & VIE2022b &   7.3 $\pm$  1.0 &    336 $\pm$      09 &          8.0 $\pm$  1.1 &    137 $\pm$      08       &    176  &      2020:049 - 2022:175 \\
NYALES20  & VIE2022b &   3.1 $\pm$  0.1 &    254 $\pm$      03 &          2.2 $\pm$  0.1 &     98 $\pm$      04       &   2282  &     1994:278 - 2022:173 \\
         & ITRF2020 &    2.3 $\pm$    0.5  &    295 $\pm$ 13  &    1.6 $\pm$    0.5  &    121 $\pm$ 19  &  &    \\
\noalign{\smallskip}\hline\noalign{\smallskip}
ONSA13SW  & VIE2022b &   3.9 $\pm$  0.3 &    207 $\pm$      05 &          1.8 $\pm$  0.3 &    177 $\pm$     10       &   85   &     2019:008 - 2022:014   \\
ONSA13NE  & VIE2022b &   5.4 $\pm$  0.3 &    213 $\pm$      03 &          2.4 $\pm$  0.3 &    179 $\pm$      07       &   104   &     2019:008 - 2022:098  \\
ONSALA60  & VIE2022b &   4.3 $\pm$  0.2 &    211 $\pm$      03 &          1.7 $\pm$  0.2 &     16 $\pm$      07       &   1162   &    1980:208 - 2022:144  \\
         & ITRF2020 &    3.5 $\pm$    0.3  &    203 $\pm$ 05  &    0.4 $\pm$    0.3  &    117 $\pm$ 49  &  &    \\
\noalign{\smallskip}\hline\noalign{\smallskip}
RAEGSMAR  & VIE2022b &   0.9 $\pm$  0.8 &    249 $\pm$     42 &          2.6 $\pm$  0.6 &     45 $\pm$     14            &    89   &     2018:352 - 2022:173   \\
         & ITRF2020 &    0.7 $\pm$    1.3  &    110 $\pm$ 106  &    0.7 $\pm$    1.3  &    116 $\pm$ 108  &  &    \\
\noalign{\smallskip}\hline\noalign{\smallskip}
RAEGYEB   & VIE2022b &   0.9 $\pm$  0.3 &    254 $\pm$     27 &          0.6 $\pm$  0.3 &     69 $\pm$     37          &    80   &    2015:048 - 2022:098  \\
YEBES40M  & VIE2022b &   5.4 $\pm$  0.3 &    242 $\pm$      03 &          3.8 $\pm$  0.3 &      6 $\pm$      04          &  391    &  2008:256 - 2022:116 \\
         & ITRF2020 &    3.4 $\pm$    0.4  &    229 $\pm$ 07  &    1.1 $\pm$    0.4  &      1 $\pm$ 23  &  &    \\
\noalign{\smallskip}\hline\noalign{\smallskip}
WETTZ13S  & VIE2022b &   4.3 $\pm$  0.3 &    227 $\pm$      04 &          0.9 $\pm$  0.3 &      4 $\pm$     17        &  88     &   2017:338 - 2022:102   \\
WETTZ13N  & VIE2022b &   6.0 $\pm$  0.3 &    243 $\pm$      03 &          1.8 $\pm$  0.3 &    127 $\pm$      09       &   418    &    2015:161 - 2022:173   \\
WETTZELL  & VIE2022b &   5.4 $\pm$  0.1 &    232 $\pm$      01 &          0.5 $\pm$  0.1 &    143 $\pm$     13       &  4080    &    1983:321 - 2022:175   \\
         & ITRF2020 &    4.1 $\pm$    0.2  &    219 $\pm$ 03  &    0.6 $\pm$    0.2  &     17 $\pm$ 22  &  &    \\
\noalign{\smallskip}\hline
\end{tabular}
\end{table*}

\begin{table}
\caption{Difference in height and in height velocity of VIE2022b w.r.t. ITRF2020 for telescopes listed in Table~\ref{tab:harmpos}.}
\label{tab:VGOSheight}       % Give a unique label
\begin{tabular}{l|r|r}
\hline\noalign{\smallskip}
antenna &   $ \Delta h$  [mm]&   $\Delta \dot{h}$ [mm/yr] \\
\noalign{\smallskip}\hline\noalign{\smallskip}
\hline
FD-VLBA  &    1.6 $\pm$    1.4 &    0.21 $\pm$   0.08 \\
GGAO12M  &    7.7 $\pm$    3.5 &   -0.28 $\pm$   0.11 \\
GGAO7108 &  -77.3 $\pm$   30.6 &   -0.28 $\pm$   0.11 \\
HRAS\_085 &   33.0 $\pm$   17.4 &    1.74 $\pm$   0.63 \\
KOKEE    &   -0.8 $\pm$    1.2 &   -0.16 $\pm$   0.08 \\
KOKEE12M &   -1.6 $\pm$    3.6 &   -0.16 $\pm$   0.08 \\
MACGO12M &   16.3 $\pm$   11.2 &    0.21 $\pm$   0.08 \\
NYALE13S &   -2.2 $\pm$   27.3 &    0.42 $\pm$   5.49 \\
NYALES20 &    4.7 $\pm$    1.1 &    0.38 $\pm$   0.09 \\
ONSA13NE &    2.6 $\pm$    3.1 &    0.01 $\pm$   0.07 \\
ONSA13SW &    4.2 $\pm$    4.0 &    0.01 $\pm$   0.07 \\
ONSALA60 &    1.4 $\pm$    1.0 &    0.01 $\pm$   0.07 \\
RAEGSMAR &  -21.7 $\pm$  390.9 &    7.52 $\pm$  96.93 \\
RAEGYEB  &    9.9 $\pm$    2.0 &   -0.43 $\pm$   0.28 \\
WETTZ13N &    3.2 $\pm$    1.1 &    0.20 $\pm$   0.06 \\
WETTZ13S &    4.0 $\pm$    1.5 &    0.20 $\pm$   0.06 \\
WETTZELL &    4.2 $\pm$    0.9 &    0.20 $\pm$   0.06 \\
YEBES40M &    2.2 $\pm$    1.3 &   -0.43 $\pm$   0.28 \\
\noalign{\smallskip}\hline
\end{tabular}
\end{table}

\begin{table*}
\caption{Transformation parameters from ITRF2020 to VIE2020 (top of the table) and from ITRF2020 to VIE2022b (bottom of the table) computed from stations with mean coordinate errors below 10~mm in Vienna TRFs. In each second row the rate per year is given.}
\label{tab:helmert}       % Give a unique label
\begin{tabular}{r|r|r|r|r|r|r}
\hline\noalign{\smallskip}
$T_x$ [mm] & $T_y$ [mm] & $T_z$ [mm] & $R_x$ [\textmu as] & $R_y$ [\textmu as]& $R_z$ [\textmu as] & scale s [ppb] ([mm])\\
$\dot{T}_x$ [mm/yr] & $\dot{T}_y$ [mm/yr] & $\dot{T}_z$ [mm/yr] & $\dot{R}_x$ [\textmu as/yr] & $\dot{R}_y$ [\textmu as/yr]& $\dot{R}_z$ [\textmu as/yr] & $\dot{s}$ [ppb/yr] ([mm/yr])\\
\noalign{\smallskip}\hline\noalign{\smallskip}
\hline
 -0.8 $\pm$   1.7  &   -3.4 $\pm$   1.7  &   -2.0 $\pm$   1.7 &    60.5 $\pm$  69.2  &   21.6 $\pm$  68.5  &   -5.4 $\pm$  54.2  &   0.56 $\pm$  0.26  (  3.6 $\pm$   1.7) \\
  0.0 $\pm$   0.0  &   -0.2 $\pm$   0.0  &    0.0 $\pm$   0.0 &     6.0 $\pm$   1.7  &    0.8 $\pm$   1.8  &    4.4 $\pm$   1.5  &   0.02 $\pm$  0.01  (  0.1 $\pm$   0.0) \\
\noalign{\smallskip}\hline\noalign{\smallskip}
  1.9 $\pm$   0.7  &   -1.0 $\pm$   0.7  &   -2.3 $\pm$   0.7 &    35.4 $\pm$  29.4  &   36.2 $\pm$  29.0  &   21.4 $\pm$  23.2  &   0.59 $\pm$  0.11  (  3.7 $\pm$   0.7) \\
  0.2 $\pm$   0.0  &   -0.2 $\pm$   0.0  &   -0.2 $\pm$   0.0 &     5.6 $\pm$   0.7  &    4.3 $\pm$   0.7  &    1.8 $\pm$   0.6  &   0.03 $\pm$  0.00  (  0.2 $\pm$   0.0) \\
\noalign{\smallskip}\hline
\end{tabular}
\end{table*}
To align the solutions on the a priori frame, no-net-translation (NNT) and no-net-rotation (NNR) conditions were applied to the coordinates and velocities of a set of 21 stable long operating telescopes:
ALGOPARK,
BR-VLBA,
FD-VLBA,
FORTLEZA,
HARTRAO,
HN-VLBA,
HOBART26,
KASHIMA,
KOKEE,
KP-VLBA,
LA-VLBA,
MATERA,
NL-VLBA,
NOTO,
NYALES20,
ONSALA60,
OV-VLBA,
SC-VLBA,
SVETLOE,
WESTFORD,
WETTZELL.
The reference epoch was set to January 1st, 2015 for both frames, VIE2020 and VIE2022b. Figure~\ref{FigTRFantennas50} shows the active telescope participation in the solution VIE2022b of telescopes which took part in more than 50 experiments as a function of time. The red crosses indicate the defined breaks in position or velocity.
\begin{figure*}
\centering
\includegraphics[width=\hsize]{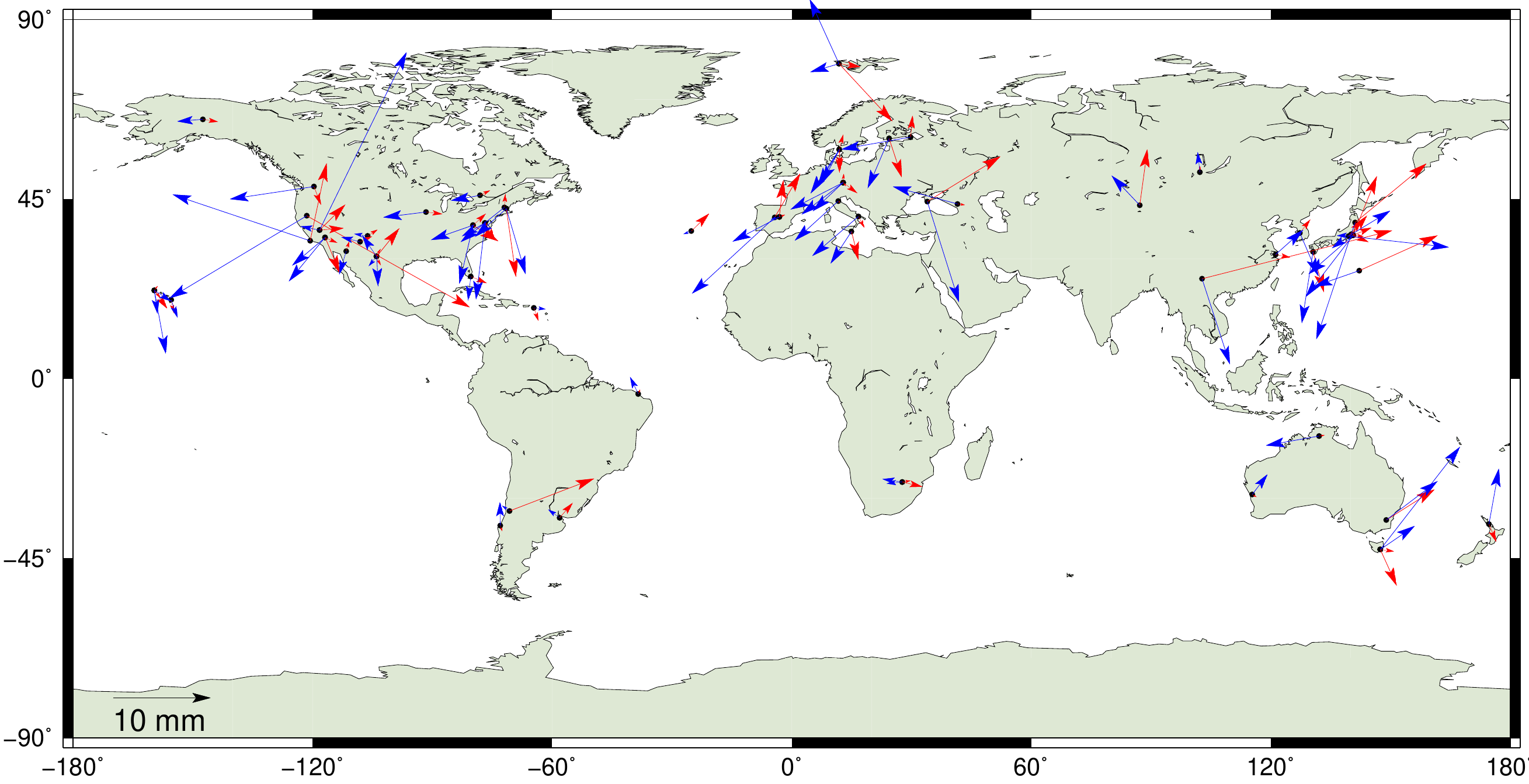}
  \caption{Annual (blue) and semi-annual (red) height signal in VIE2022b. The amplitude is depicted by the arrow length. The phase of the maximum displacement is represented by the orientation of the arrow. If the arrow points towards north, the maximum appears in January and it continues clockwise further. The maximum of the semi-annual signal is depicted in the first half year.}
     \label{Fig:TRFharmonics}
\end{figure*}
\begin{figure}
   \centering
   \includegraphics[width=\hsize]{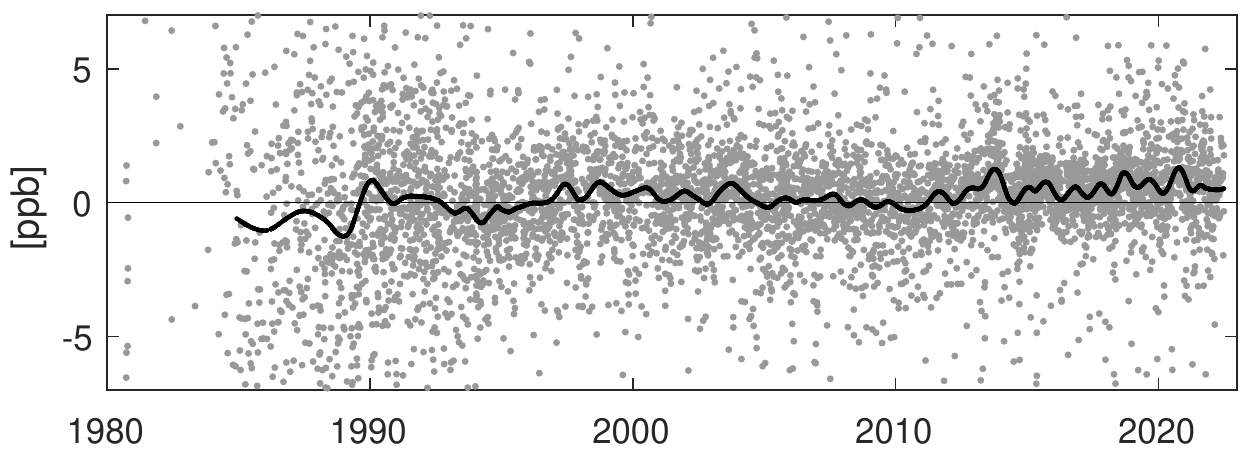}
      \caption{Session-wise scale factor computed with VIE2022b parametrization w.r.t. ITRF2020. The black line represents the smoothed scale factor obtained by local regression using a span of 5\% of the total number of data points.}
         \label{FigTRF_scale}
\end{figure}

In addition, the seasonal displacement in the height component at annual and semi-annual periods $P_i$ ($P_1 = 365.25$ days, $P_2 = 182.63$ days) at chosen stations were estimated. The station height displacement $\Delta h = \Delta h_1 + \Delta h_2$ is parametrized in a form of sine and cosine amplitude ($A_{si}, A_{ci}$):
   \begin{equation}
\Delta h_i =
A_{ci} \cdot \cos\bigg(\frac{mjd-\mathrm{mjd_0}}{ P_i }\cdot2\pi\bigg) + A_{si} \cdot \sin\bigg(\frac{mjd-\mathrm{mjd_0}}{ P_i }\cdot2\pi\bigg)
,
   \end{equation}
where the reference epoch $\mathrm{mjd_0}$ was set to January 1st, 2000 and $mjd$ is the modified Julian date of the experiment. The amplitude $A_i$ and phase $\phi_i$ of the seasonal wave is obtained with the basic mathematical relation as:
\begin{equation}
A_i = \sqrt{A_{ci}^2  + A_{si}^2 },\qquad
\phi_i = \arctan{\bigg(\frac{A_{si}}{A_{ci}} \bigg)}.
\end{equation}
For these parameters, we selected 74 antennas which observed in at least 50 experiments distributed uniformly over a year. To estimate the seasonal variation of the signal, the measurements have to cover the changing seasons. For this reason, we did not estimate
the seasonal displacement for EFLSBERG, OHIGGINS, PARKES, SYOWA and YLOW7296 because their participation in VLBI sessions is rather episodic (compare with Fig.~\ref{FigTRFantennas50}). The estimation of the amplitude for a semi-annual signal is relevant especially for telescopes in the subtropical zones where the sum of the annual and semi-annual signal fits the changes in the telescope height over a year best. The estimated annual and semi-annual signals for all 74 telescopes are plotted in Fig.~\ref{Fig:TRFharmonics} as arrows representing the amplitude (length) and phase (direction, azimuth 0 = Jan 1) of the height displacement at the respective periods. In Table~\ref{tab:harmpos}, we highlight the estimates of the harmonic signal at the newly established VGOS antennas and complete the information with the co-located legacy telescopes. The accuracy of the estimated signal depends on the number and distribution of the experiments allowing the correct tracing of the height change over the year. For example, in Europe, the maximum reading of the annual signal in the height occurs in the summer months July and August, which means that the crust goes up. At observatories in Onsala or Wettzell, there is an agreement between the harmonic height changes estimated at all telescopes with a positive maximum of 4-6~mm occurring in July. This can be explained by the minimal water content in the ground and no snow load since we applied neither the harmonic signal provided for ITRF2014 or ITRF2020 nor any hydrology loading modeling in the data analysis. These results agree with former findings published, e.g., by \citet{Tesmer2009} or \citet{Krasna2015}. For comparison, Table~\ref{tab:harmpos} includes the amplitude and phase for annual and semi-annual signals in height computed from the sine and cosine amplitudes as they are provided with ITRF2020\footnote{\href{ https://itrf.ign.fr/ftp/pub/itrf/itrf2020/ITRF2020-Frequencies-ENU-CF.snx}{ITRF2020-Frequencies-ENU-CF.snx}}. In ITRF2020 an identical signal at all telescopes within each geodetic site is given. We would like to point out, that for telescope NYALE13S an erroneous harmonic signal of zero amplitude is given in ITRF2020, which should be treated carefully in a potential VLBI analysis.

To complete the information about the difference in height at VGOS antennas from Table~\ref{tab:harmpos} in VIE2022b and ITRF2020, the height differences for the epoch 2015 and the differences in height velocity computed as VIE2022b minus ITRF2020 are listed in Table~\ref{tab:VGOSheight}. The formal errors are computed as propagation of uncertainties of both catalogs. At most telescopes the estimated heights are higher in VIE2022b compared to ITRF2020. In particular, it is true for all differences exceeding their formal error and telescopes observing in the final years. Differences in the estimated height which lie above the three-fold formal error occur at NYALES20 (4.7 $\pm$ 1.1~mm), RAEGYEB (9.9 $\pm$ 2.0~mm) and WETTZELL (4.2 $\pm$ 0.9~mm). This agrees with the fact that the scale factor of the VIE2022b is higher than the scale factor of ITRF2020. Table~\ref{tab:helmert} summarizes 14 transformation parameters (three translation parameters ($T_x$, $T_y$, $T_z$), three rotation parameters ($R_x$, $R_y$, $R_z$) and one scale factor with their time derivatives) computed from stations with mean coordinate error $m_{xyz}$ (Eq.~(\ref{eq_mce})) lower than 10~mm in VIE2020 and VIE2022b, respectively:
\begin{equation}\label{eq_mce}
     m_{xyz} = \sqrt{\big(m_{x}^2 + m_{y}^2 + m_{z}^2\big)/3},
\end{equation}
where $m_{x}^2, m_{y}^2,$ and $m_{z}^2$ are formal errors of the coordinates.
The top lines of the table include parameters describing the transformation from ITRF2020 to VIE2020, and the bottom part lists parameters from ITRF2020 to VIE2022b. The maximum translation of $-3.4 \pm 1.7$~mm occurs in y-direction between VIE2020 and ITRF2020 but it is still within the three-fold of its formal error. The remaining translation parameters are negligible. All rotations are close or below their formal errors and the maximum value of $60.5 \pm 69.2~$\textmu as  appears in x-direction between VIE2020 and ITRF2020. The scale factors from the Vienna VLBI TRFs VIE2020 and VIE2022b show a difference of $0.56 \pm 0.26$~ppb and $0.59 \pm 0.11$~ppb w.r.t. ITRF2020, respectively. The ITRF2020 relies by its scale definition on two space geodetic techniques, namely VLBI and SLR. The scale of the ITRF2020 long-term frame was determined using internal constraints in such a way that there are zero scale and scale rate between ITRF2020 and the scale and scale rate averages of VLBI selected sessions up to 2013.75 (see ITRF2020 webpage\footnote{\href{https://itrf.ign.fr/en/solutions/ITRF2020}{https://itrf.ign.fr/en/solutions/ITRF2020}}). After 2014 there is an increase in the scale determined from the VLBI technique w.r.t. ITRF2020 as can be seen in Fig.~\ref{FigTRF_scale}.

%   \begin{figure}
%   \centering
%   \includegraphics[width=\hsize]{dWRMS_onlyVIEglob2020_morethan2mm.pdf}
%      \caption{              }
%         \label{xx}
%   \end{figure}

%__________________________________________________________________

\section{Celestial reference frame}
\label{sec_crf}
Celestial reference frames (CRF) are estimated in common global solutions together with the terrestrial reference frames. We focus on the CRF including the most recent geodetic and astrometric VLBI experiments until June 2022 observed at S/X frequencies. The catalog VIE2022b-sx consists of 5407 radio sources. All sources present in the underlying data are kept in the analysis and their positions are estimated as global parameters. This means that we neither remove gravitational lenses from the solution, as it was done in ICRF3, nor do we estimate sources with non-linear motions as session-wise parameters (so-called special handling sources in ICRF2~\citep{Fey2015}). We use ICRF3 as a priori frame and model the galactic acceleration correction with the adopted ICRF3 value of $A_G = 5.8~$\textmu as/yr for the amplitude of the solar system barycenter acceleration vector in direction to the Galactic center (right ascension $\alpha_{GC} = 266.4^{\circ}$, declination $\delta_{GC} =-28.94^{\circ}$) for the epoch 2015.0.

\begin{figure}
\centering
\includegraphics[width=\hsize]{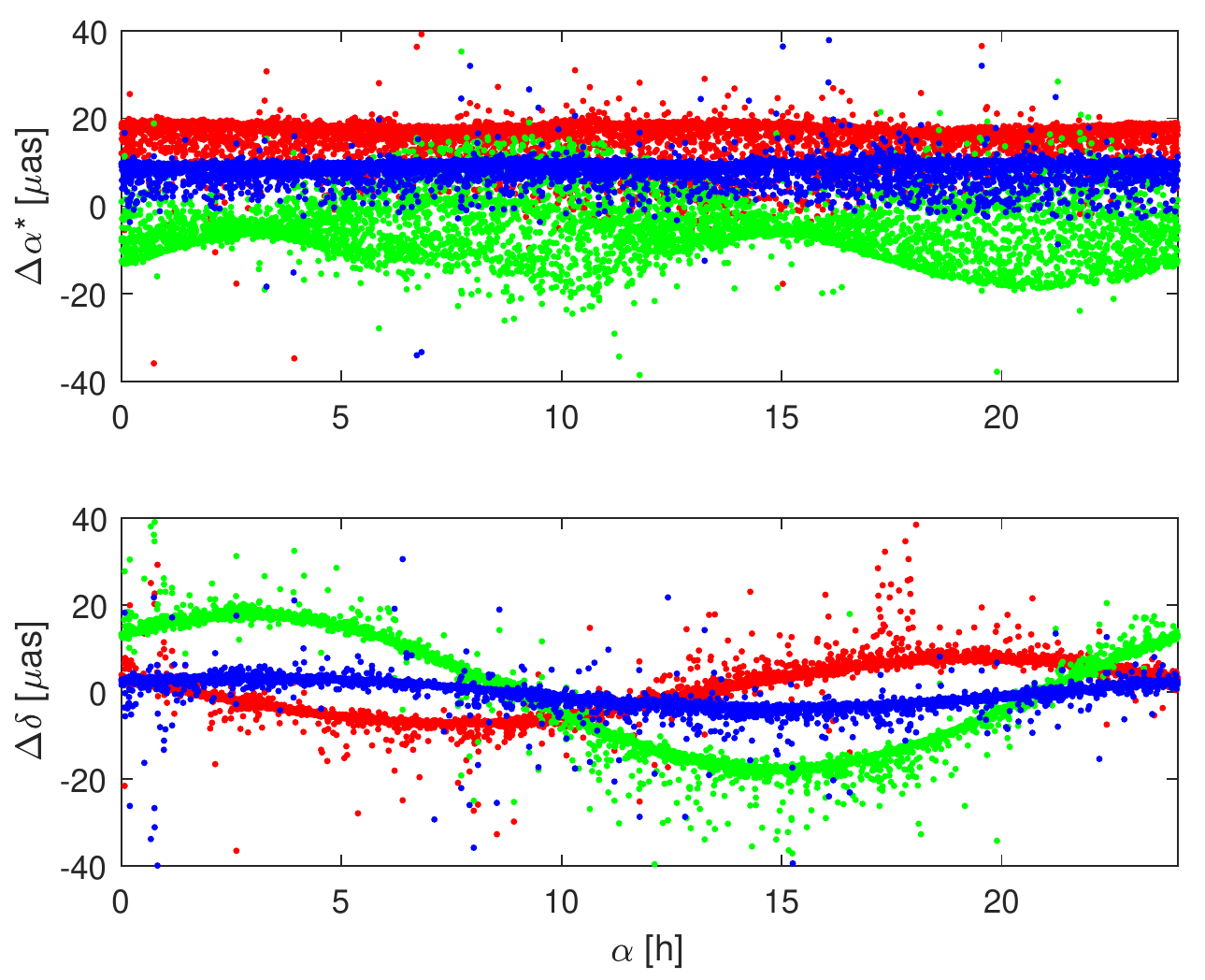}
  \caption{Radio source coordinate estimates from three solutions applying the NNR condition on three different subgroups of 101 ICRF3 defining sources each. The differences are plotted w.r.t. a solution aligned with all 303 ICRF3 defining sources.}
     \label{FigCRF_diffDATUM_wrtRA}
\end{figure}

The alignment of the frame is accomplished by unweighted no-net-rotation constraints \citep[NNR;][]{Jacobs2010} to the positions of 301 ICRF3 defining sources. In ICRF3, a new set of defining sources was selected (independent of ICRF2 defining sources) based on several criteria, one of them being a uniform distribution on the sky. Because of the generally sparser distribution of the observed radio sources in the far south, sources with lower position stability or a low number of observations had to be included in the set of defining sources. However, after several tests, we excluded two ICRF3 defining sources (0700-465 and 0809-493) from the NNR condition since their observations led to a large distortion of the newly estimated frame. In Fig.~\ref{FigCRF_diffDATUM_wrtRA}, we show results of one of our tests regarding the ICRF3 datum sources. We divided the 303 ICRF3 datum sources into three groups so that we included every third source from the ICRF3 datum source list sorted in ascending order according to their right ascension. Then we computed three global solutions which we aligned to ICRF3 with NNR applied to coordinates of the 101 ICRF3 datum sources included in the three independent lists. Fig.~\ref{FigCRF_diffDATUM_wrtRA} shows the estimated source coordinates $\alpha^*$ and $\delta$ with respect to a solution where all 303 ICRF3 datum sources were in the NNR condition. We use the designation $\alpha^*$ for right ascension scaled by declination of the source, i.e., $\alpha^* = \alpha \cdot \cos{\delta}$. Each color represents one of the global solutions aligned with the subgroup of 101 ICRF3 defining sources. A systematic difference in the estimated coordinates when excluding 202 radio sources from the NNR condition is evident with the peak of the differences at around 20~\textmu as for both coordinates. In particular, we identified two radio sources with a difference in the estimated coordinates above 1~mas when dropped from the NNR condition (0700-465, $\Delta \alpha^* =3.9$~mas, $\Delta \delta = 3.5$~mas; 0809-493, $\Delta \alpha^* = 4.2$~mas, $\Delta \delta = 1.7$~mas). Therefore we did not use them to align our Vienna celestial reference frames (described below) to ICRF3. With more observations of these sources, their position variability can be studied in the future.

%\begin{figure}
%\centering
%\includegraphics[width=\hsize]{vie2022b_diffDATUM_wrtDe.pdf}
%  \caption{}
%     \label{FigCRF_diffDATUM_wrtDe}
%\end{figure}

\begin{figure}
\centering
\includegraphics[width=\hsize]{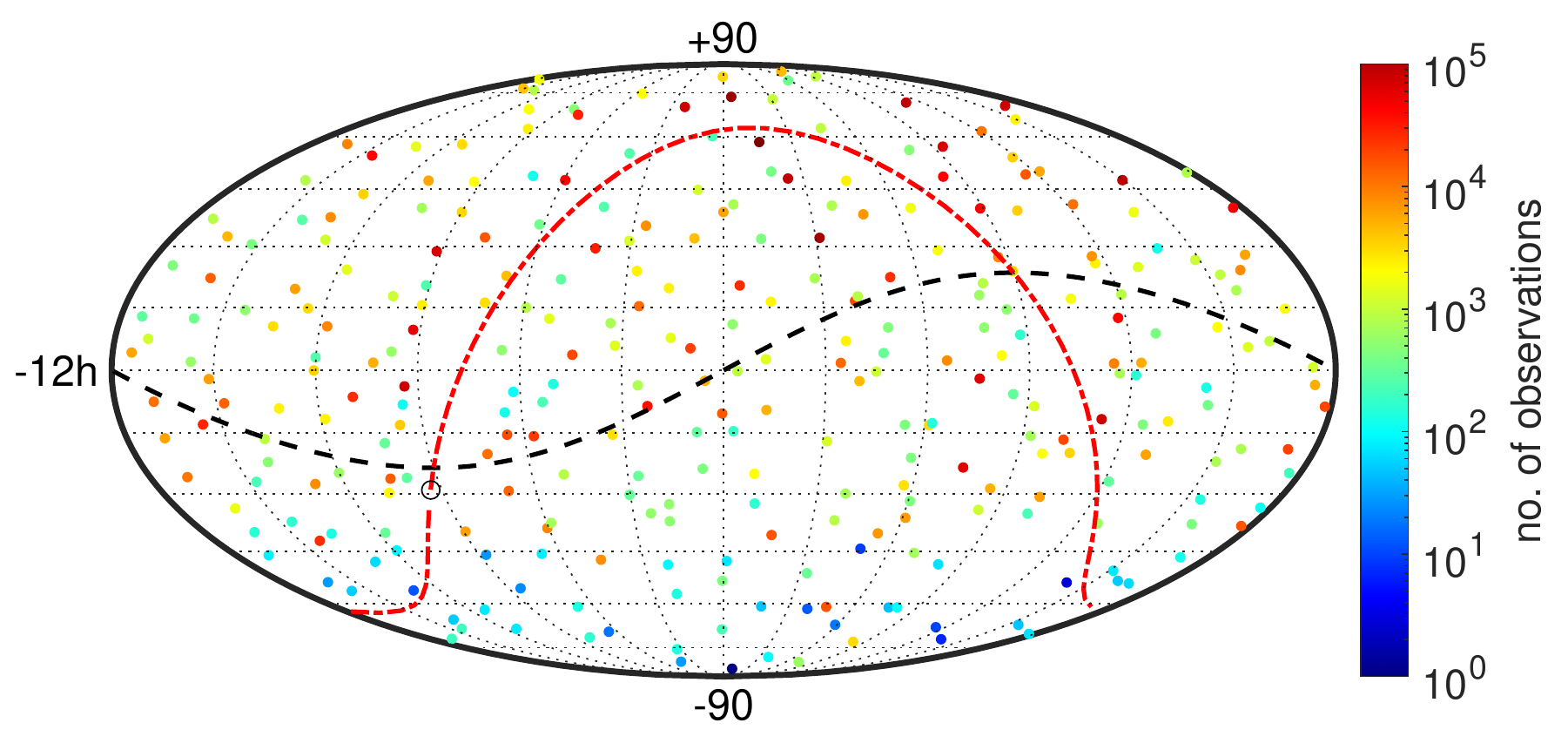}
  \caption{Increase in number of observations for the ICRF3 defining sources by 2022.5 after the release of ICRF3. The black dashed line depicts the ecliptic plane and the red line represents the galactic plane with the Galactic center denoted as an empty black circle.}
     \label{FigCRF_obsdiff_vie2022b_icrf3D}
\end{figure}
\begin{figure}
\centering
\includegraphics[width=\hsize]{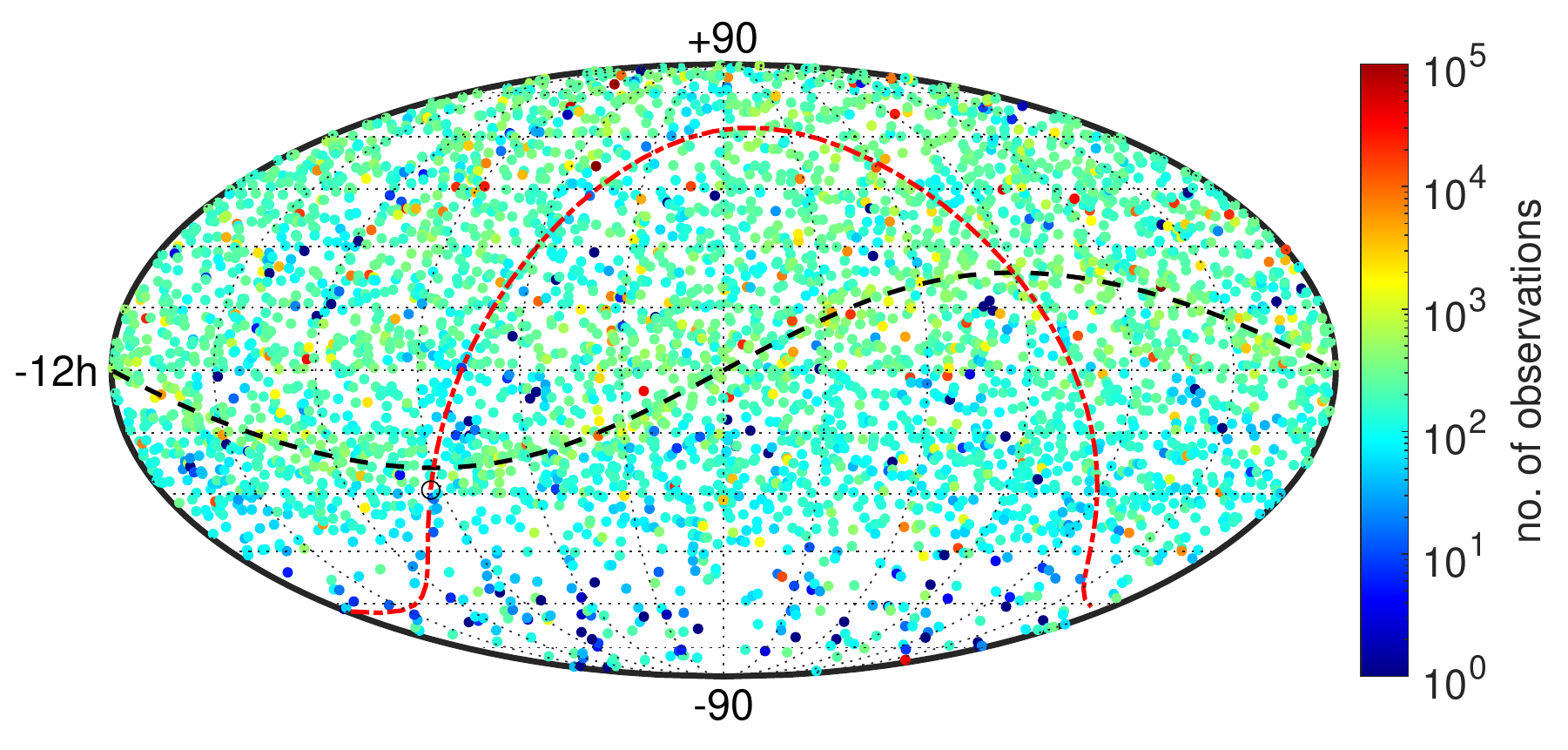}
  \caption{Increase in number of observations for the ICRF3 non-defining sources by 2022.5 after the release of ICRF3.}
     \label{FigCRF_obsdiff_vie2022b_icrf3noD}
\end{figure}
\begin{figure}
\centering
\includegraphics[width=\hsize]{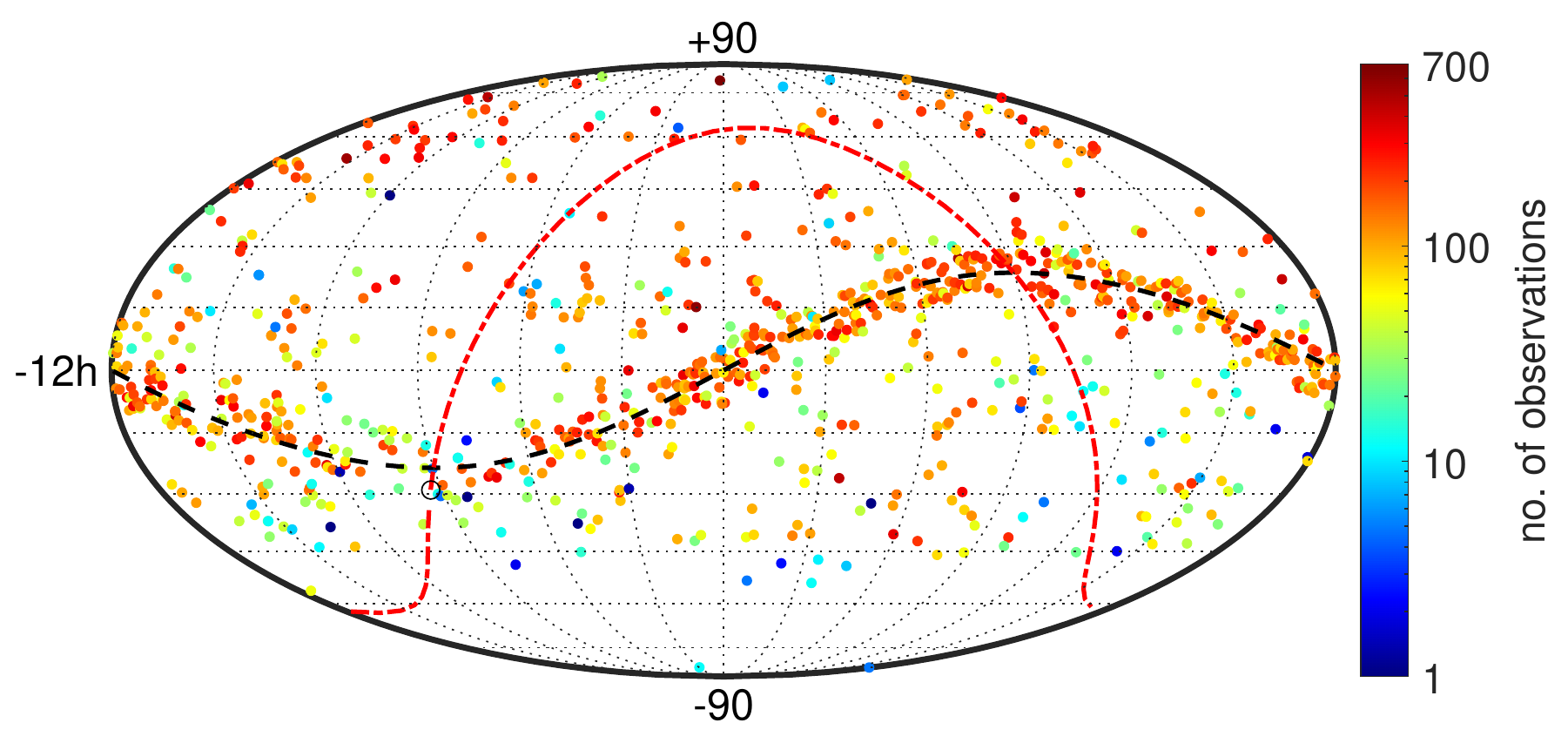}
  \caption{Number of observations for new sources after the release of ICRF3 until 2022.5.}
     \label{FigCRF_obsdiff_vie2022b_NOTINicrf3}
\end{figure}
We depict the number of new observations for ICRF3 defining sources in Fig.~\ref{FigCRF_obsdiff_vie2022b_icrf3D} and non-defining sources in Fig.~\ref{FigCRF_obsdiff_vie2022b_icrf3noD} after the ICRF3 cutoff date June 2022. The majority of new observations for the deep south sources ($\delta<-45^\circ$) come from three dedicated programs: astrometric IVS Celestial Reference Frame Deep South sessions \citep[IVS-CRDS; ][]{deWitt2019}, sessions performed under the umbrella of the Asia-Oceania VLBI Group for Geodesy and Astrometry \citep[AOV; ][]{McCallum2019}, and sessions conducted with the geodetic Australian mixed-mode program \citep{McCallum2022}. In addition, Fig.~\ref{FigCRF_obsdiff_vie2022b_NOTINicrf3} shows the 870 sources which were observed in geodetic/astrometric VLBI experiments after the ICRF3 release for the first time. The majority of the new sources was observed in dedicated astrometry experiments conducted by the AOV and by the Very Long Baseline Array \citep[VLBA; ][]{vlba1995}. In particular, the goal was to increase the number of sources at the ecliptic plane available for the navigation of interplanetary spacecrafts \citep{deWitt2022}. Since the VLBA is located on United States territory, the network is able to observe radio sources only down to approximately -45$^{\circ}$ declination. Table~\ref{tab:souObsStat} summarizes the statistics on observations from the Figs.~\ref{FigCRF_obsdiff_vie2022b_icrf3D}~-~\ref{FigCRF_obsdiff_vie2022b_NOTINicrf3}. It is evident that despite the great international effort \citep{deWitt2021}, the increase of observations to southern sources is still slower than the increase to northern sources, mainly due to the lack of large geodetic VLBI dishes in the Southern Hemisphere needed for observations of faint sources.
\begin{table}
\caption{Statistics on observations in VIE2022b-sx since the ICRF3 cutoff date. The values are divided for ICRF3 defining sources (def), ICRF3 non-defining sources (non-def) and sources not included in ICRF3 (new). The number of sources, number of observations and a median value for observations per source is given separately for three declination zones separated by $\delta = 0^\circ$ and $\delta = -45^\circ$.}
\label{tab:souObsStat}       % Give a unique label
\begin{tabular}{l|c|r|r|r}
\hline\noalign{\smallskip}
  & $\delta$& no. of  & no. of & median of  \\
    &&  sources & obs. & obs. per sou.  \\
\noalign{\smallskip}\hline\noalign{\smallskip}
\hline
def    & <0$^{\circ}$, 90$^{\circ}$>  & 149& $1.6\cdot10^6$  &	3395 \\
      &  <-45$^{\circ}$, 0$^{\circ}$> & 105& $7.3\cdot10^5$  &	2111 \\
       &<-90$^{\circ}$, -45$^{\circ}$>  & 46& $4.5\cdot10^4$ &	273 \\
       \hline
non-def  & <0$^{\circ}$, 90$^{\circ}$> 	&  2467     &   $1.5\cdot10^6$ &  205  \\
&  <-45$^{\circ}$, 0$^{\circ}$> &1561 &  $4.6\cdot10^5$ & 125\\
 &<-90$^{\circ}$, -45$^{\circ}$>  &209 &   $5.7\cdot10^4$ & 62\\
 \hline
new   &  <0$^{\circ}$, 90$^{\circ}$> &    493 &    $1.0\cdot10^5$   &    198   \\
&  <-45$^{\circ}$, 0$^{\circ}$> &364 & $5.6\cdot10^4$ &140 \\
& <-90$^{\circ}$, -45$^{\circ}$> &13 &$5.4\cdot10^2$ &28 \\
\noalign{\smallskip}\hline
\end{tabular}
\end{table}

\begin{figure}
   \centering
   \includegraphics[width=\hsize]{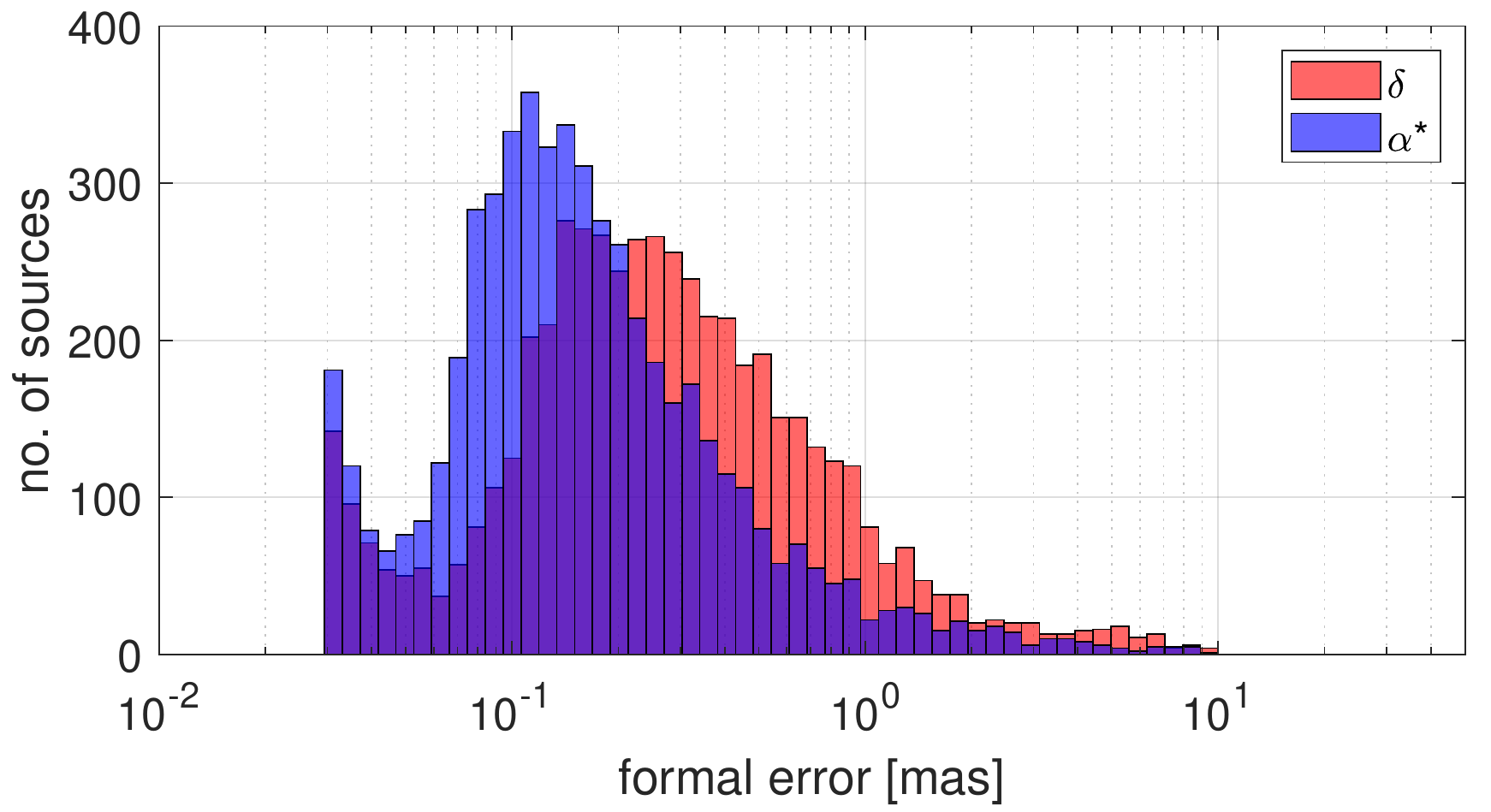}
      \caption{Distribution of source coordinate formal errors in VIE2022b-sx.}
         \label{FigCRF_formal_errors_histogram_vie2022bsx}
\end{figure}

\begin{figure}
    \centering
    \includegraphics[width=\hsize]{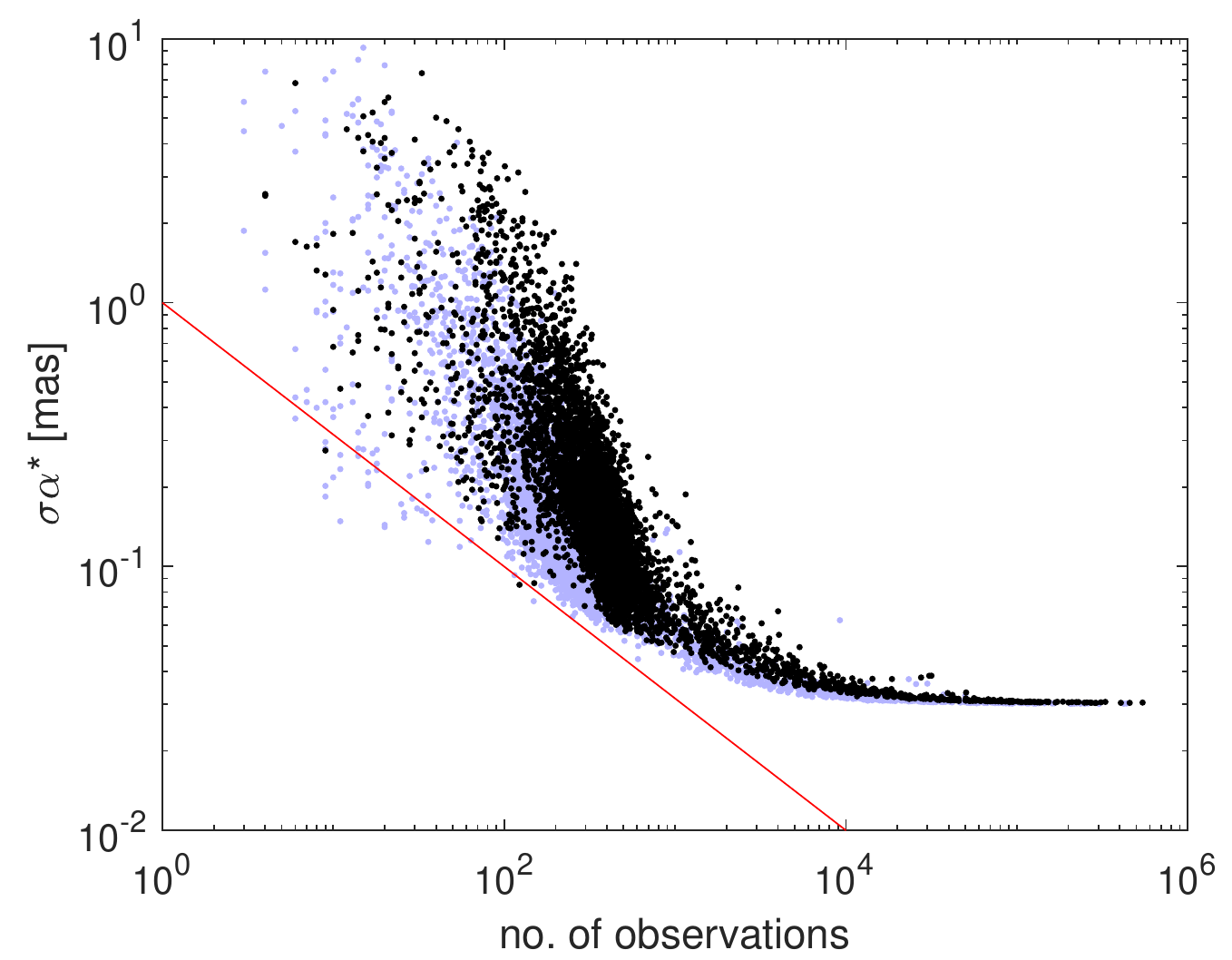}\\
    \includegraphics[width=\hsize]{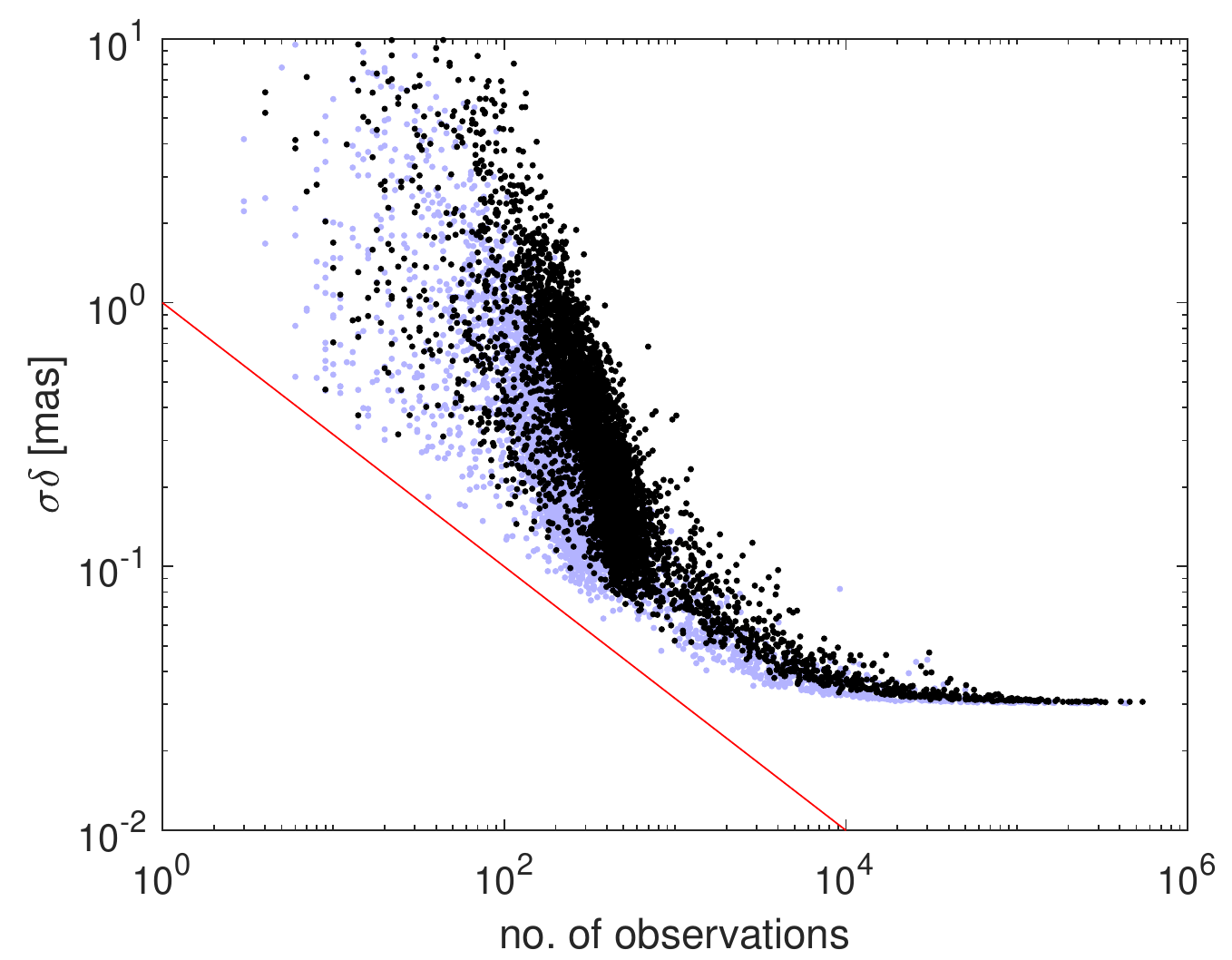}
      \caption{Formal errors in $\alpha^*$ (upper plot) and $\delta$ (lower plot) with respect to number of observations in VIE2022b-sx (black dots) and ICRF3 (light blue dots). The red line depicts the hypothetical decrease of formal errors with the square root of the number of observations.}
         \label{FigCRF_vie2022b_sigma_numobs_log}
\end{figure}

The histogram of VIE2022b-sx formal errors (Fig.~\ref{FigCRF_formal_errors_histogram_vie2022bsx}) gives an overview of the distribution of uncertainties in $\alpha^*$ and $\delta$. The median formal error computed over all sources is 143~\textmu as for $\alpha^*$ and 250~\textmu as for $\delta$ corresponding to the peaks of the histogram. The ratio between the median formal error for $RA^*$ and $De$ of a factor of two remains similar to that in ICRF3 (i.e., 127~\textmu as / 218~\textmu as). In the VIE2022b-sx CRF catalog, we follow the recommendation of the ICRF3 which advises inflating the formal errors of the source coordinates from the global least squares adjustment by a factor of 1.5 and then adding a noise floor of 30~\textmu as in quadrature to avoid the dropping of the uncertainties to unrealistic values for frequently observed sources \citep{Charlot20}.
The individual formal errors of source coordinates in VIE2022b-sx (black dots) and in ICRF3 (light blue dots) with respect to the number of observations are shown in Fig.~\ref{FigCRF_vie2022b_sigma_numobs_log}. Theoretically, if there were no correlations between individual observations, the estimated formal errors would drop with the square root of observations, represented as the red line in Fig.~\ref{FigCRF_vie2022b_sigma_numobs_log}. The deviation from this rule, as can be seen in the figure, is caused by accounting of the elevation dependent weighting of observations in VIE2022b-sx which changes the stochastic model and by applying the noise floor which influences the formal errors of the most frequently observed sources.


\begin{figure}
   \centering
   \includegraphics[width=\hsize]{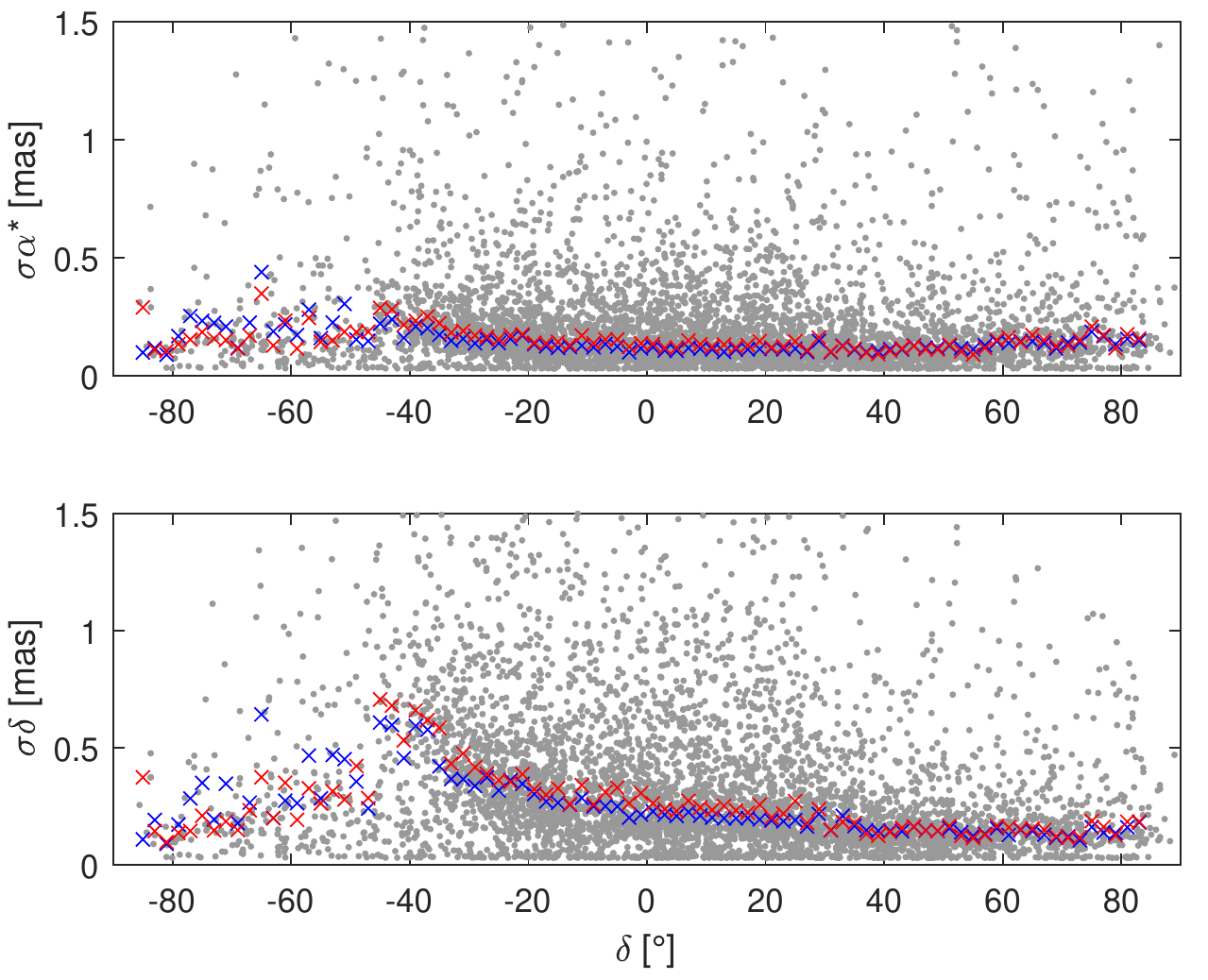}
      \caption{VIE2022b-sx formal errors (grey dots) in $\alpha^*$ (upper plot) and $\delta$ (lower plot) with respect to declination. Crosses depict median formal errors computed over 2$^{\circ}$ declination in red for VIE2022b-sx and in blue for ICRF3.}
         \label{FigCRF_vie2022b_dec_DesigmaRAcosdec}
\end{figure}

Fig.~\ref{FigCRF_vie2022b_dec_DesigmaRAcosdec} shows the formal errors of VIE2022b-sx source coordinates (grey dots) with respect to declination. The red crosses depict the median formal error computed over 2$^{\circ}$ wide declination zones. Especially uncertainties in declination are showing a growth starting at 30$^{\circ}$ declination (median $\sigma De$ = 150~\textmu as) which accelerates until $-45^{\circ}$ declination (median $\sigma De$ = 700~\textmu as). Further south, the formal errors jump back to values around 250~\textmu as. The explanation can be found in the station network. The majority of sources was observed in campaigns of Very Long Baseline Array Calibrator Survey \citep[VCS;][]{Beasley02, Fomalont2003, Petrov2005, Petrov2006, Petrov2008, Kovalev2007} or VCS-II \citep{Gordon2016} conducted by the VLBA network which, based on its location, observes the southern sources under rather low elevation angles. This means that the path of the signal in the atmosphere is longer, which leads to larger formal errors of the estimated position of the emitting radio sources.

\begin{figure}
   \centering
   \includegraphics[width=\hsize]{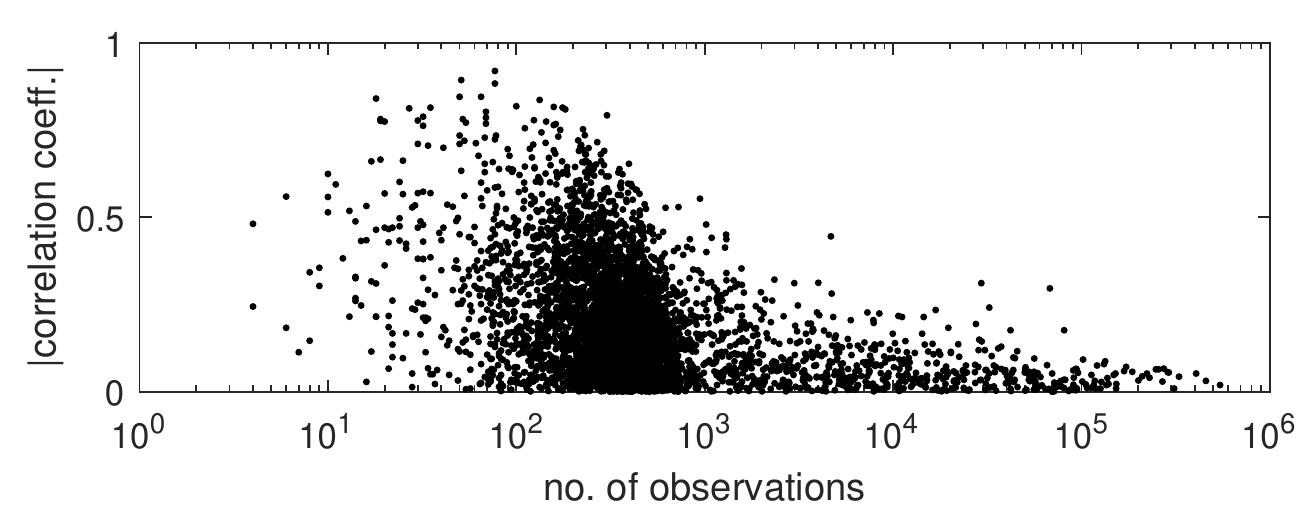}
      \caption{Correlation coefficients between $\alpha^*$ and $\delta$ with respect to number of observations in VIE2022b-sx.}
         \label{FigCRF_vie2022b_correl_numobs}
   \end{figure}

In Fig.~\ref{FigCRF_vie2022b_correl_numobs}, we plot the correlation coefficient between $\alpha^*$ and $\delta$ in VIE2022b-sx with respect to the number of observations of the respective source. The median absolute correlation coefficient is 0.15 which implies a weak correlation between the two coordinates. Nevertheless, the plot shows that the correlation for sources with a lower number of observations can be strong and the correlation coefficient can be close to 1. With an increasing number of observations, the maximal possible correlation between the two estimated source coordinates decreases to about 0.3 for $10^3$ observations and stays below 0.1 for the most observed sources.

The difference between VIE2022b-sx and ICRF3 is further analyzed with a vector spherical harmonics decomposition \citep[VSH; ][]{Mignard2012, Titov2013, mayer2020}. Table~\ref{tab:souVSH} shows
16 estimated parameters relevant to the second degree VSH, i.e., rotation ($R_1, R_2, R_3$), dipole ($D_1, D_2, D_3$), and ten coefficients ($a$) for the quadrupole harmonics of magnetic ($m$) and electric ($e$) type. Prior to the comparison, outliers from the VIE2020b-sx CRF were removed. As outliers, we defined sources with an angular separation to ICRF3 larger than 10~mas or with a position formal error higher than 10~mas. In total, 46 outliers were removed from the VSH computation. The list of sources with angular separation to ICRF3 > 10~mas is given in Table~\ref{tab:souout}. The non-linear change in position after the ICRF3 release for some of them (3C48, CTA21, 1328+254) was reported by e.g., \citet{Frey2021} or \citet{Titov2022}. The causes for the position change of the remaining sources are subject to further investigation. \\
The VSH between two catalogs ($cat1$ and $cat2$) is obtained with a least squares adjustment for the common radio sources. Unlike the NNR condition in the Vienna global solution, the least squares adjustment for the VSH estimation is carried out using a weight matrix $P$, where $P = P_{cat1} + P_{cat2}$ and the $P_{cat}$ contains correlation between $\alpha^*$ and $\delta$ for the respective source as reported in the catalog. The elements of $P_{cat}$ for one source $i$ are computed as:
\begin{equation}
  \begin{aligned}
 & P_{i\alpha^*, i\alpha^*} = (\sigma_\alpha \cdot \cos \delta)^{-2},\\
 & P_{i\delta, i\delta} = (\sigma_\delta )^{-2},\\
 & P_{i\alpha^*, i\delta} = P_{i\delta, i\alpha^*} = (cr \cdot \sigma_\alpha \cdot \cos \delta \cdot \sigma_\delta )^{-1},
  \end{aligned}
\end{equation}
where $cr$ stands for the correlation coefficient. In the first two columns of Table~\ref{tab:souVSH}, we show the weighted VSH computed between VIE2022b-sx and ICRF3, and between VIE2022b-sx and USNO-2022July03\footnote{\href{ https://crf.usno.navy.mil/quarterly-vlbi-solution}{https://crf.usno.navy.mil/quarterly-vlbi-solution}}. USNO-2022July03 is a CRF solution computed in the same manner as the ICRF3 by the USNO where the time span of processed VLBI experiments is similar to VIE2022b-sx. For both CRF catalogs, ICRF3 and USNO-2022July03, the difference to VIE2022b-sx is described with similar VSH estimates where each of them is below the noise floor of 30~\textmu as of the catalogs. The maximum rotation parameter is 13~\textmu as for $R_3$ between VIE2022b-sx and ICRF3. The two largest VSH parameters are the $D_3$ and the quadrupole term $a_{2,0}^{e}$ which reach $-25 \pm  2~$\textmu as and $20 \pm  3~$\textmu as for VIE2022b-sx w.r.t. ICRF3, and  $-18 \pm  1~$\textmu as and $11 \pm  2~$\textmu as for VIE2022b-sx w.r.t. USNO-2022July03. The connection of these two parameters to the modeling of tropospheric gradients in the VLBI analysis and their constraints applied in the adjustment was shown, e.g., by \citet{mayer2020}. Additionally, we computed the VSH between VIE2022b-sx and USNO-2022July03 without the weight matrix $P$. In the last column of Table~\ref{tab:souVSH} one sees a slight decrease of the $D_3$ and $a_{2,0}^{e}$ parameters. One possible explanation lies in the larger declination formal errors of sources located around $-40^{\circ}$ declination as discussed in Fig~\ref{FigCRF_vie2022b_dec_DesigmaRAcosdec}. Taking their formal errors as weights into account in the VSH computation can lead to the larger distortion in the north-south direction described by the $a_{2,0}^{e}$ term.

\begin{table}
\caption{VSH parameters up to degree two for VIE2022b-sx minus ICRF3, and VIE2022b-sx minus USNO-2022July03 after eliminating outliers. The parameter "weights" denotes the weight matrix applied in the VSH least squares adjustment. Units are \textmu as.}
\label{tab:souVSH}       % Give a unique label
\begin{tabular}{l|r|r|r}
\hline\noalign{\smallskip}
  & \multicolumn{3}{|c}{VIE2022b-sx versus}   \\
           &  ICRF3&\multicolumn{2}{|c}{USNO-2022July03}    \\
weights &yes& yes & no\\
\noalign{\smallskip}\hline\noalign{\smallskip}
\hline
$R_1$ &  $ +9 \pm  2$ 	 &  $+11 \pm  1$ 	&  $ +3 \pm  6$ 		 \\
$R_2$ &  $ -9 \pm  2$ 	 &  $ +4 \pm  1$ 	&  $-11 \pm  6$ 		 \\
$R_3$ &  $-13 \pm  2$ 	 &  $ -7 \pm  1$ 	&  $-10 \pm  6$ 		 \\
\hline
$D_1$ &  $ +1 \pm  2$ 	 &  $ -2 \pm  1$ 	&  $-14 \pm  6$ 		 \\
$D_2$ &  $ -6 \pm  2$ 	 &  $ -7 \pm  1$ 	&  $ -7 \pm  6$ 		 \\
$D_3$ &  $-25 \pm  2$ 	 &  $-18 \pm  1$ 	&  $-14 \pm  6$ 		 \\
\hline
$a_{2,0}^{e}$  & $+20 \pm  3$ 	& $+11 \pm  2$ 	& $ -8 \pm  7$ 	 \\
$a_{2,0}^{m}$  & $ +6 \pm  2$ 	& $ -1 \pm  1$ 	& $-10 \pm  7$ 	 \\
$a_{2,1}^{e,Re}$ & $ -3 \pm  3$& $ -2 \pm  2$ 	& $-10 \pm  8$ 	 \\
$a_{2,1}^{e,Im}$ & $ +1 \pm  3$& $ -0 \pm  2$ 	& $ -2 \pm  8$ 	 \\
$a_{2,1}^{m,Re}$ & $ -2 \pm  3$& $ -0 \pm  2$ 	& $ +4 \pm  8$ 	 \\
$a_{2,1}^{m,Im}$ & $ +3 \pm  3$& $ +5 \pm  2$ 	& $ -6 \pm  8$ 	 \\
$a_{2,2}^{e,Re}$ & $ -0 \pm  1$& $ +1 \pm  1$ 	& $ -5 \pm  4$ 	 \\
$a_{2,2}^{e,Im}$ & $ -4 \pm  1$& $ -3 \pm  1$ 	& $ +3 \pm  4$ 	 \\
$a_{2,2}^{m,Re}$ & $ +4 \pm  1$& $ -0 \pm  1$ 	& $ +0 \pm  4$ 	 \\
$a_{2,2}^{m,Im}$ & $ +1 \pm  1$& $ -0 \pm  1$ 	& $ -4 \pm  4$ 	 \\
\noalign{\smallskip}\hline
\end{tabular}
\end{table}

%\begin{table*}
%\caption{VSH parameters up to degree two between VIE2022b-sx and ICRF3 after outlier elimination.}
%\label{tab:souVSH}       % Give a unique label
%\begin{tabular}{r|r|r|r|r|r|r|r|r|r}
%$R_1$ &$R_2$ &$R_3$ &$D_1$ &$D_2$ & $D_3$ & & & & \\
%$+20 \pm  2$ & $-31 \pm  2$& $-20 \pm  2$ &	 $ +4 \pm  2$& $ -6 \pm  2$ & $ -3 \pm  2$ & & & & \\
%$a_{2,0}^{e}$  & $a_{2,0}^{m}$  &$a_{2,1}^{e,Re}$ &$a_{2,1}^{e,Im}$ & $a_{2,1}^{m,Re}$ & 	$a_{2,1}^{m,Im}$ %&$a_{2,2}^{e,Re}$ & $a_{2,2}^{e,Im}$ &$a_{2,2}^{m,Re}$ &$a_{2,2}^{m,Im}$ \\
%$+25 \pm  2$ & $ +5 \pm  2$ & $ -2 \pm  2$& $ +1 \pm  3$& $ -2 \pm  2$& $ +4 \pm  2$ &$ -0 \pm  1$ & $ -4 \pm % 1$ 	& $ +3 \pm  1$&$ +2 \pm  1$\\
%\end{tabular}
%\end{table*}

 \begin{table*}
\caption{List of sources with angular separation between ICRF3 and VIE2022b-sx larger than 10~mas.}
\label{tab:souout}       % Give a unique label
\begin{tabular}{l|l|r|r|r|r|r|r|r}
\hline\noalign{\smallskip}
IERS  & IVS & $\Delta\alpha^*$  & $\Delta\delta$ &angular separation & first obs. & last obs. & no. of  & no. of\\
name  & name  &  [mas] & [mas] &  [mas]&[mjd] & [mjd] &  sessions & obs. \\
\noalign{\smallskip}\hline\noalign{\smallskip}
\hline
0106-391 & -         &  $    -3.51 \pm   4.80$ & $    -23.60 \pm  18.08$ &  $    23.86 \pm  17.89$ & 58203.3  & 59460.3  &      6 &     64 \\
0134+329 & 3C48      &  $     1.25 \pm   0.05$ & $    -56.85 \pm   0.08$ &  $    56.87 \pm   0.08$ & 48193.8  & 59378.0  &     49 &   1736 \\
0201-440 & -         &  $     1.34 \pm  17.40$ & $    -99.25 \pm  55.92$ &  $    99.26 \pm  55.92$ & 58143.4  & 59508.8  &      4 &     15 \\
0316+162 & CTA21     &  $     2.10 \pm   0.06$ & $    -10.22 \pm   0.12$ &  $    10.44 \pm   0.12$ & 50084.5  & 59378.0  &     17 &   1299 \\
0328-060 & -         &  $    29.73 \pm   4.54$ & $    -16.02 \pm   6.97$ &  $    33.77 \pm   5.19$ & 56874.5  & 59440.3  &      8 &     54 \\
0350+177 & -         &  $    -6.78 \pm   0.84$ & $     63.43 \pm   1.33$ &  $    63.79 \pm   1.33$ & 57924.7  & 59405.2  &      6 &    116 \\
0512-129 & -         &  $    -3.68 \pm   1.74$ & $      9.31 \pm   4.41$ &  $    10.01 \pm   4.15$ & 58143.4  & 59522.9  &      5 &     69 \\
0709+008 & -         &  $     7.36 \pm   2.74$ & $      7.27 \pm   3.12$ &  $    10.35 \pm   2.94$ & 52939.7  & 58631.3  &      7 &     82 \\
0748-378 & -         &  $    -9.33 \pm  10.52$ & $     48.22 \pm  25.45$ &  $    49.12 \pm  25.07$ & 57011.1  & 59508.8  &      8 &     40 \\
0753-425 & -         &  $     1.46 \pm   0.73$ & $     12.36 \pm   2.24$ &  $    12.45 \pm   2.22$ & 55370.8  & 59522.9  &      7 &    123 \\
0903-392 & -         &  $     1.93 \pm   4.72$ & $    -15.35 \pm  14.04$ &  $    15.47 \pm  13.94$ & 57046.0  & 58981.5  &      7 &     32 \\
0932-281 & -         &  $     6.54 \pm   1.79$ & $      7.87 \pm   4.55$ &  $    10.23 \pm   3.68$ & 50687.3  & 59508.8  &      6 &     99 \\
0951+699 & -         &  $    12.00 \pm  35.25$ & $     -4.94 \pm  34.56$ &  $    12.98 \pm  35.15$ & 58203.3  & 58592.8  &      3 &     12 \\
1015-314 & -         &  $     3.58 \pm   2.21$ & $    -17.51 \pm   5.26$ &  $    17.87 \pm   5.17$ & 52305.8  & 59560.6  &      8 &     77 \\
1117-248 & -         &  $   -12.40 \pm   2.21$ & $     11.22 \pm   3.09$ &  $    16.72 \pm   2.64$ & 50631.3  & 59463.5  &     12 &     71 \\
1306+660 & -         &  $   -15.08 \pm   3.58$ & $    -33.07 \pm   4.77$ &  $    36.35 \pm   4.59$ & 57011.1  & 59405.2  &      8 &     65 \\
1305-241 & -         &  $     6.90 \pm   8.14$ & $     14.74 \pm   9.89$ &  $    16.27 \pm   9.60$ & 58158.9  & 59440.3  &      5 &     44 \\
1328+254 & -         &  $     8.48 \pm   0.57$ & $     17.13 \pm   0.89$ &  $    19.11 \pm   0.83$ & 52408.7  & 58644.9  &      6 &    164 \\
1422+268 & -         &  $    -2.98 \pm   4.87$ & $    -12.57 \pm   4.64$ &  $    12.91 \pm   4.66$ & 58136.6  & 58981.5  &      4 &     46 \\
1507-246 & -         &  $    70.00 \pm   1.80$ & $   -128.92 \pm   3.36$ &  $   146.70 \pm   3.08$ & 57924.7  & 59611.7  &      8 &     68 \\
1539-093 & -         &  $   -29.52 \pm  12.78$ & $     13.61 \pm  10.61$ &  $    32.50 \pm  12.43$ & 50575.3  & 58981.5  &      9 &     36 \\
1612+797 & -         &  $     7.06 \pm   0.64$ & $     -7.36 \pm   0.74$ &  $    10.20 \pm   0.70$ & 53780.1  & 58510.3  &      6 &    237 \\
1657-298 & -         &  $   346.60 \pm   5.03$ & $   -687.18 \pm   8.32$ &  $   769.64 \pm   7.76$ & 57973.7  & 59611.7  &      7 &     40 \\
1706-223 & -         &  $    -3.66 \pm   0.64$ & $    -14.04 \pm   1.73$ &  $    14.51 \pm   1.68$ & 57011.1  & 58746.6  &      5 &    123 \\
1711-251 & -         &  $   213.09 \pm 188.33$ & $   -466.99 \pm 364.28$ &  $   513.31 \pm 340.50$ & 57596.8  & 58981.5  &      7 &     16 \\
1755+626 & -         &  $   -21.04 \pm   2.94$ & $    -41.25 \pm   2.63$ &  $    46.31 \pm   2.70$ & 55370.8  & 59522.9  &      9 &    105 \\
1829-106 & -         &  $    21.41 \pm   4.07$ & $    -35.84 \pm   3.56$ &  $    41.74 \pm   3.70$ & 51731.8  & 59560.6  &     10 &     17 \\
1858-143 & -         &  $    -2.82 \pm  12.25$ & $     28.09 \pm  16.33$ &  $    28.23 \pm  16.29$ & 58203.3  & 58981.5  &      4 &     23 \\
1934-638 & -         &  $   -22.59 \pm   0.88$ & $      2.69 \pm   0.72$ &  $    22.75 \pm   0.88$ & 48765.9  & 59065.7  &      8 &     36 \\
2028-204 & -         &  $   494.59 \pm  15.25$ & $  -1021.10 \pm  32.61$ &  $  1134.58 \pm  30.10$ & 58203.3  & 59460.3  &      5 &     19 \\
2105-212 & -         &  $     9.91 \pm   1.25$ & $     -4.23 \pm   2.42$ &  $    10.77 \pm   1.49$ & 57011.1  & 59535.8  &      8 &     83 \\
2216-007 & -         &  $    73.13 \pm   2.36$ & $    -85.80 \pm   3.11$ &  $   112.73 \pm   2.82$ & 56266.8  & 58644.9  &      6 &     80 \\
2219-340 & -         &  $    13.60 \pm   6.72$ & $     10.41 \pm  18.62$ &  $    17.13 \pm  12.51$ & 57098.3  & 58981.5  &      7 &     36 \\
2318-195 & -         &  $    10.27 \pm   0.73$ & $     20.12 \pm   1.80$ &  $    22.59 \pm   1.64$ & 58143.4  & 59460.3  &      6 &    101 \\
2346+750 & -         &  $    -1.74 \pm   0.47$ & $     10.99 \pm   0.63$ &  $    11.12 \pm   0.62$ & 57808.9  & 59560.6  &      7 &    134 \\
\noalign{\smallskip}\hline
\end{tabular}
\end{table*}

\section{Summary}

We presented high-quality celestial and terrestrial reference frames as provided by the IVS Special Analysis Center VIE. We compared our recent solution which includes IVS sessions until June 2022 with the current international reference frames, i.e., ICRF3 and ITRF2020, and highlighted the differences coming from the additional VLBI observations after the data deadlines for the international frames, i.e., spring 2018 and December 2020, respectively. The consistent Vienna celestial and terrestrial reference frames were estimated in a common global least squares adjustment and are publicly available together with the corresponding Earth orientation parameters at \href{ https://www.vlbi.at/index.php/products}{https://www.vlbi.at}. In this paper, we provided a detailed description of the underlying a priori models and the selected parametrization for the global least squares solution.

\begin{acknowledgements}
The authors acknowledge the International VLBI Service for Geodesy and Astrometry (IVS) and all its components for providing VLBI data. J.B. and F.J. would like to thank the Austrian Science Fund (FWF) for supporting their work with project VGOS-Squared (P~31625).
\end{acknowledgements}

% for the bibliography, at the end
\bibliographystyle{aa} % style aa.bst
\bibliography{reference_krasna} % your references Yourfile.bib

\end{document}